\documentclass[preprint2]{aastex63}
\usepackage{lineno}

\newcommand{\kms}{{km s$^{-1}$}}
\newcommand{\kmsMpc}{{km s$^{-1}$ Mpc$^{-1}$}}

\newcommand{\myemail}{mglee@astro.snu.ac.kr}

\usepackage{amsmath}
\usepackage{multirow}
\hypersetup{linkcolor=red,citecolor=blue,filecolor=cyan,urlcolor=magenta}

\accepted{ApJL}

\shorttitle{SMACS J0723.3-7327}
\shortauthors{Lee et al.}
\begin{document}

\title{
Detection of Intracluster Globular Clusters in the First JWST Images of the Gravitational Lens Cluster SMACS J0723.3-7327 at $z=0.39$}

\correspondingauthor{Myung Gyoon Lee}
\email{\myemail} 
\author[0000-0003-2713-6744]{Myung Gyoon Lee}
\author[0000-0002-4477-1208]{Jang Ho Bae}
\affil{Astronomy Program, Department of Physics and Astronomy, SNUARC, Seoul National University, 1 Gwanak-ro, Gwanak-gu, Seoul 08826, Republic of Korea}

\author[0000-0002-2502-0070]{In Sung Jang}
\affiliation{Department of Astronomy \& Astrophysics, University of Chicago, 5640 South Ellis Avenue, Chicago, IL 60637, USA}

\begin{abstract}
We present a survey of globular clusters (GCs) in the massive gravitational lens cluster SMACS J0723.3-7327 at $z=0.39$ based on the early released JWST/NIRCam images.
In the color-magnitude diagrams of the point sources we find clearly a rich population of intracluster GCs that spread in a wide area of the cluster. 
Their ages, considering the cluster redshift, are younger than 9.5 Gyr. The F200W (AB) magnitudes of these GCs, $26.5<{F200W_0} <29.5$ mag, correspond to $-15.2<{M_{F200W}} <-12.2$ mag, showing that they belong to the brightest GCs (including ultracompact dwarfs). 
The spatial distributions of these GCs  show a megaparsec-scale structure elongated along the major axis of the brightest cluster galaxy. In addition, they show a large number of substructures, some of which are consistent with the substructures seen in the map of diffuse intracluster light. 
The GC number density map is, in general, consistent with the dark matter mass density map based on the strong lensing analysis in the literature.
The radial number density profile of the GCs in the outer region is steeper  than the dark matter mass profile obtained from lensing models.
These results are consistent with those for the GCs found in the deep HST images of Abell 2744, another massive cluster at $z=0.308$, and in simulated galaxy clusters.
This shows that the intracluster GCs are an excellent independent tool to probe the dark matter distribution in galaxy clusters as well as to reveal the cluster assembly history in the JWST era. 
\end{abstract}


\section{Introduction}

Massive galaxy clusters that show (strong and weak) gravitational lensing effects are ideal targets to investigate the large scale distribution of dark matter  and the relation between the dark matter and the baryonic matter (e.g., \citep{lot17,fox22} and references therein). The baryonic matter in galaxy clusters are composed mainly of galaxies and intracluster medium (ICM). Primary components in the ICM are X-ray emitting hot gas, intracluster light (ICL), intracluster globular clusters (GCs), and intracluster planetary nebulae. Among them, the intracluster GCs are, because they are compact and abundant, very useful to study the relation between dark matter and baryonic matter at large scales. 

Wide field imaging has been a main tool for the survey of intracluster GCs in the nearby galaxy clusters \citep{lee10,dur14,pen11, iod17,har17, hil18, mad18, har20}.
High resolution power of the HST enables us to probe the intracluster GCs in the more distant galaxy clusters, and study the relation between the intracluster GC  and dark matter distributions \citep{ala13,ala17,lee16a}.

Using the deep ACS/F814W and WFC3/F105W images of the HST Frontier Fields \citep{lot17}, \citet{lee16a} presented a survey of GCs including ultracompact dwarfs (UCDs) in the massive merging cluster Abell 2744 (known as the Pandora cluster) at
$z=0.308$. They found a rich population of GCs and UCDs in the cluster, a significant fraction of which are intracluster GCs that spread in the wide area of the cluster. They showed that the number density map of the intracluster GCs in Abell 2744 is similar to the dark matter mass map derived from CATS strong lensing models \citep{jau15}.

In this study, we search for GCs of SMACS J0723.3-7327 (called SMACS 0723 hereafter), a massive cluster at $z=0.39$, which is farther than Abell 2744,  using the first  JWST/NIRCam images released on July 12, 2022 as part of the early released JWST images \citep{pon22,rig22}. 
Recently \citet{mow22} found five GCs that are considered to be associated with the Sparkler galaxy at $z=1.38$ which is gravitationally lensed by SMACS 0723 from the JWST NIRCam images and HST/ACS images.
Our primary goal is to find and study the GCs that belong to SMACS 0723. 
Basic properties of SMACS 0723 are listed in {\bf Table \ref{tab:info}}.

HST observations of this cluster were included in the HST Snapshot survey \citep{rep18} and the RELICS program \citet{coe19},
providing the magnitudes of the brightest cluster galaxy (BCG),  F814W = 19.14 ($M_{F814W} = -22.9$) mag 
and F606W=20.35 ($M_{F606W}=-22.0$) mag.
We checked first the HST data of SMACS 0723 available in the archive, but found that the images are too shallow to detect GCs in this cluster. The JWST/NIRCam data go much deeper for these GCs which are composed of old stars, enabling us to detect them.
 SMACS 0723 shows a large number of gravitationally lensed images of background sources, which 
 were used for tracing dark matter in the cluster \citep{pas22,gol22, cam22,mah22}. 
 The mass of SMACS 0723 based on strong lensing models is $M (R<400 {\rm kpc})\sim 3 \times 10^{14} M_\odot$ \citep{mah22}, and the mass derived from PLANCK observations is $M({\rm PLANCK})=8.39 \times 10^{14} M_\odot$ \citep{pas22}. 
\citet{mah22} presented a mean redshift of 26 cluster members located in the central region of SMAC 0723, $z=0.3877$,  which is slightly smaller than the redshift of the BCG, $z=0.3912$. In the following analysis, we adopt $z=0.39$ and the position of the BCG as the cluster center. 
We adopt the cosmological parameters: 
$H_0=$ 67.8 km s$^{-1}$ Mpc$^{-1}$, $\Omega_{\Lambda}=0.7$, and $\Omega_{\Lambda}=0.3$.
At the adopted distance of SMACS 0723 ($d_L=2170$ Mpc, $(m-M)_0=41.69$), one arcsec (arcmin) corresponds to 5.453 (327.16) kpc.

This paper is organized as follows. Data and data reduction are described in Section 2, and main results  are given in Section 3. The main results are discussed in Section 4, and are summarized in the final section.

\section{Data}

In the data set of the first released JWST images of SMACS 0723, we use NIRCam images obtained with
three short wavelength (SW) filters (F090W, F150W, F200W) and three long wavelength (LW) filters (F277W, F356W, F444W) on June 7, 2022 (see the field locations in Figure 1 of \citet{pon22}). The total exposure time is 7537 sec for each filter.
The SW filter images have an image scale of $0\farcs03$ per pixel, and the LW images have 
a twice larger image scale of $0\farcs06$ per pixel.

\begin{figure*}
    \centering
\includegraphics[scale=0.25]{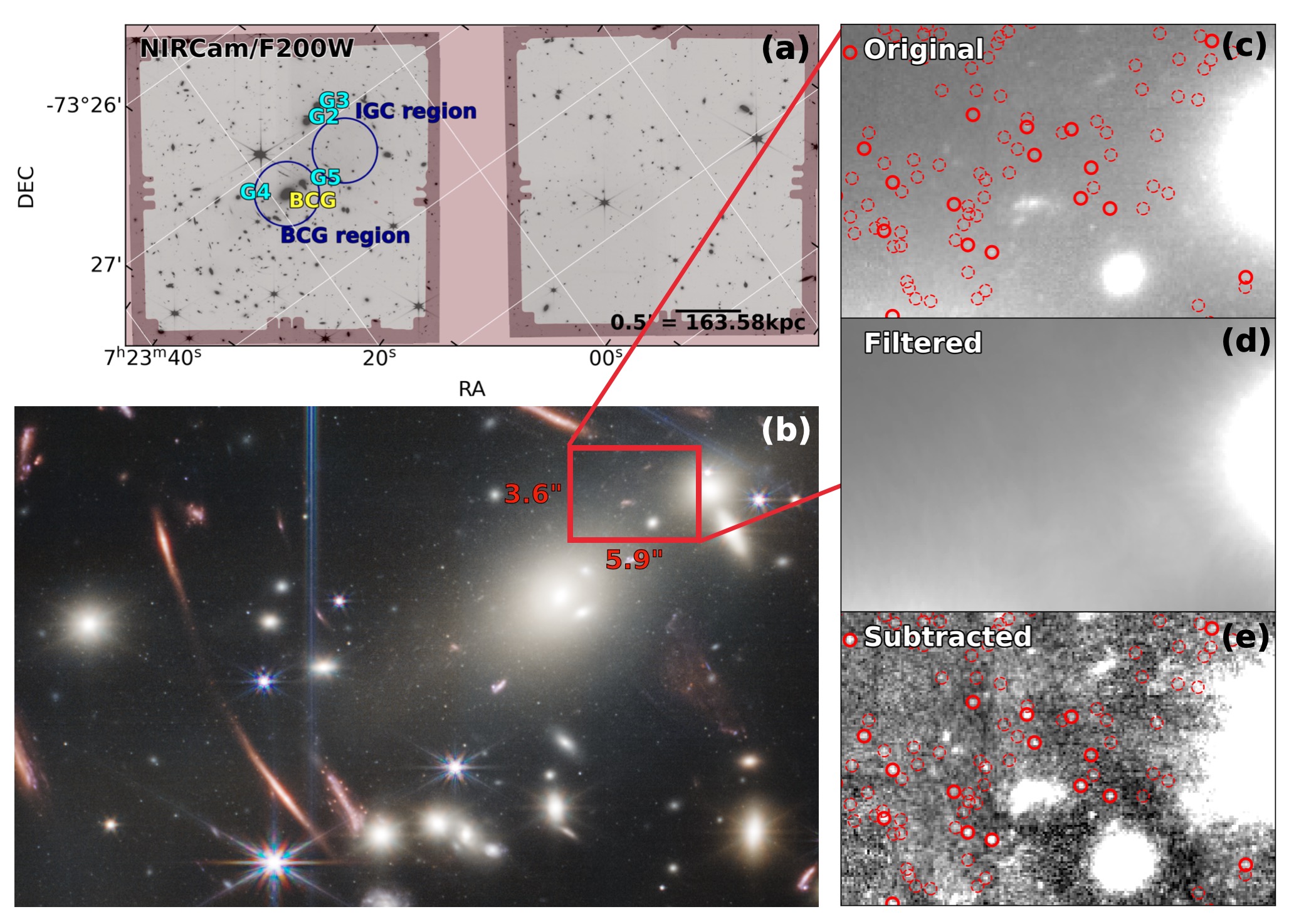} 
    \caption{(a) Gray scale maps of the F200W images of the central (left) and parallel (right) fields of SMACS 0723 (see also Figure 1 of \citet{pon22}). Two large circles with radius of $15\arcsec$  represent the BCG region and intracluster GC region, and the brightest galaxies including the BCG are labeled.
    Burgundy color denotes the edge regions with low signal to noise ratios in the combined images, which were not used for analysis.
    (b) A zoomed-in view of the BCG region in  the central field of SMACS 0723: the Sparkler galaxy ($z=1.38$, with a red linear shape embedding several red point sources) is  at
    $15\arcsec$ east of the BCG.
    (c, d, and e) Zoomed-in views of the original F200W image of the red box region in (b), the same image smoothed with a ring median filter, and the background-subtracted image. Red circles represent the point sources (thin dotted line) and GC candidates (thick solid line)  found in this study, which are visible more clearly in the background-subtracted image.}
    
	\label{fig:finder}
\end{figure*}

The released images include two fields: a cluster central field that covers the central region of the cluster, and a parallel field that covers the outskirt of the cluster, as shown in  {\bf Figure \ref{fig:finder}(a)}. 
The field of view for each of the NIRCam images is $129\arcsec \times 129\arcsec$, and the gap between the central field and the parallel field is $44\arcsec$.  The parallel field is $\sim2\farcm9$  to the southwest from the central field. The radial coverage of the parallel field is 550 kpc $<R<$ 1.3 Mpc, where $R$ is the projected radial distance from the cluster center.
In {\bf Figure \ref{fig:finder}(a)} we display  gray scale maps of F200W images of the two fields.
We mark the brightest galaxies (G2, G3, G4, and G5) including the BCG in the central field. 
The edge regions of each field marked by burgundy color have low signal-to-noise ratios in the combined images so they are not used for the final analysis.
In {\bf Figure \ref{fig:finder}(b,c,d, and e)}
we also display the zoomed-in color image of the central field, showing the BCG region,  and the zoomed-in images of the BCG region that show GCs clearly, which will be described in the following.

\section{Data Reduction}

We use F200W images, which have the highest sensitivity among the six NIRCam images, as a reference image for point source detection\footnote{JWST user documentation, https://jwst-docs.stsci.edu}.
We adopt the AB magnitudes in this study. The relations between the AB and Vega magnitudes in the 
JWST CRDS\footnote{https://jwst-crds.stsci.edu} are:
mag(AB)--mag(Vega)=0.504, 1.243, 1.706, 2.315, 2.824, and 3.242 for F090W, F150W, F200W, F277W, F365W, and F444W.

Mean effective radii of typical globular clusters in the Milky Way are about 3 pc and the FWHM values of the point sources in the JWST/NIRCam/SW images are about 2.4 pixels
($0\farcs072$=0.39 kpc).
Thus typical GCs appear as  point sources in the JWST/NIRCam images of SMACS 0723 (see also \citet{fai22}).
We select the point sources from which GCs are segregated as follows.

\subsection{Photometry of the Extended Sources}

First we obtain the photometry of the sources (including both extended and point sources) detected with 10$\sigma$ threshold in the images  using Source Extractor \citep{ber96}. We carry out dual mode photometry with 
the F277W image 
as a reference image for source detection.
As the pixel scales for LW and SW detector are different, the SW images were reprojected and aligned with respect to the F277W image using the wregister task in IRAF  \citep{tod86, sts12} and astroalign \citep{ber20}, respectively. For aligning the images, we use the positions of several stars.
We use LW images for the analysis of galaxies because the normal red sequence of galaxies in galaxy clusters is better distinguishable in the LW color-magnitude diagrams (CMDs), than in the SW CMDs, as shown in the following.

\begin{figure*}
    \centering
\includegraphics[scale=0.2]{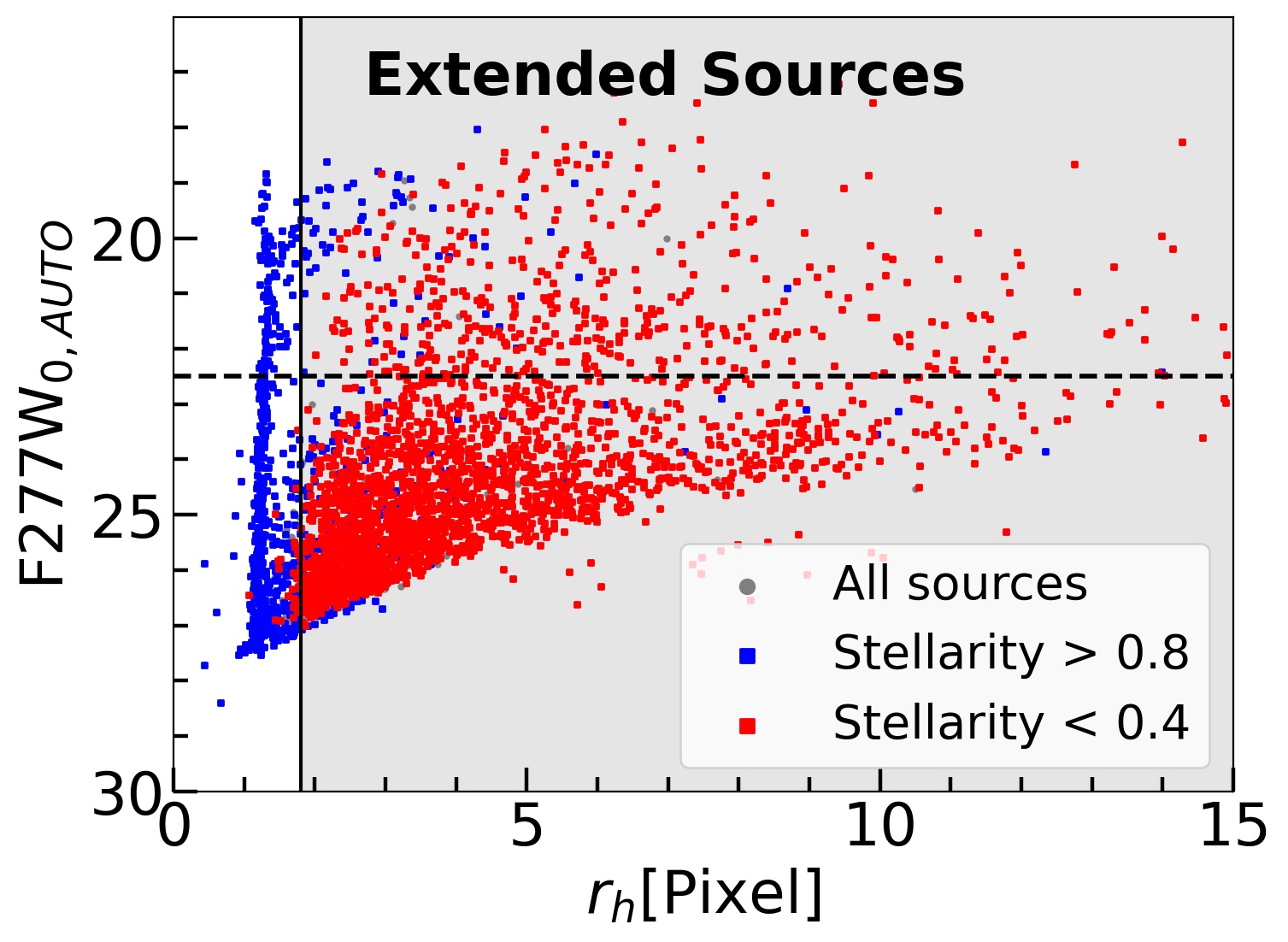}
    \caption{Effective radius $r_h$ vs. 
    F277W (AUTO) magnitudes of the sources
    detected with the Source Extractor. Red circles represent the  sources with stellarity $<0.4$, and blue circles denote the sources with stellarity$ >0.8$.
    The narrow slanted line dominated by the blue circles represent the point sources. 
    Note that some bright sources ($F277W_0<22.5$ mag) with stellarity$>0.8$ have large effective radii, $r_h >1.8$ pixels. 
    In this study we select  extended sources using the effective radius criterion:  
    $r_h >1.8$ pixels in the F277W image, as marked by the vertical black line.}
	\label{fig:rhmag}
\end{figure*}

{\bf Figure \ref{fig:rhmag}} display effective radius $r_h$ vs. 
F277W (AUTO) magnitudes of the sources     detected with the Source Extractor. 
    We also mark the  sources with small stellarity $<0.4$ (red circles), and  the  sources with large stellarity $ >0.8$ (blue circles). 
    The prominent narrow slanted sequence dominated by blue circles represent the point sources with small effective radii. 
    Some of the bright sources with large effective radii are found to have a large value of stellarity $>0.8$ so we consider that effective radius is a better indicator to distinguish extended sources than stellarity. 
    In this study we select  extended sources using the effective radius criterion:  
    $r_h >1.8$ pixels in the F277W image. 

The extended sources include galaxies and gravitationally-lensed sources.
We adopt AUTO magnitudes for the extended sources provided by the Source Extractor, and use them for the analysis of photometric properties of the galaxies in the following. 

\subsection{Photometry of the Point Sources}

For better detection of the point sources in the images that include a number of bright galaxies, we subtract the diffuse background light from the original images. 
First, we derive a smooth background image of the original F200W image, using a ring median filter with $r_{in}=18 $ pixel and $r_{out}=20$ pixel.
Then we subtract this smooth background image from the original image, and use the background-subtracted image for the point source detection.

\subsubsection{Point Spread Function Fitting Photometry} 

We obtain the point spread function (PSF) fitting photometry of the sources in the F150W and F200W images of the central and parallel fields using DAOPHOT \citep{ste87}.
We derive the PSF using the isolated bright stars in the images.
Among the detected sources we select the point sources  using the sharpness parameter and photometric errors provided by DAOPHOT. 

We calibrate the instrumental PSF-fitting magnitudes
using the total magnitudes of bright stars  given in the source catalog of SMACS 0723 included in the released data.
We use only the bright stars  $21<F150W, F200W<24$ mag which have small errors. 

However, it is noted that this zero-point calibration is preliminary. There could be a substantial (systematic) error associated with the pre-flight version of the JWST catalog \citep{boy22}.
The catalog of the point source photometry in this study will be publicly available at GitHub\footnote{https://github.com/hanlbomi/SMACS-J0723.3-7327-GCs}.

We calculate the extinction values for the JWST filters from the extinction value for $V$-band given by \citet{sch98,sch11} using the extinction law  \citep{fit99,ind05} and the filter information  in the SVO filter profile service website.\citep{rod12, rod20}. 
The extinction values are listed in Table \ref{tab:info}. 
We apply the foreground extinction correction to the apparent magnitudes, 
use the subscript 0 for the foreground-extinction-corrected magnitudes.

\subsubsection{Completeness Tests}

We estimate the photometric completeness of the point sources using the artificial star experiments with DAOPHOT/ADDSTAR. We added ~2500 artificial stars with a range of magnitude and color to each of F150W and F200W images. We repeat this process 400 times for each set of images so that the total number of added artificial stars in each band is one million. We obtain the photometry of these sources using the same procedure as used for the original images. 
Then we select the recovered/added stars with the GC color range, and estimate the completeness from the number ratio of the recovered stars to the added stars as a function of magnitude. Figure 3 displays the completeness vs. magnitude for F150W and F200W bands. We fit the data using the interpolation function given in \citet{har09}: 
   $ f(m) = 0.5 (\beta-\alpha (m-m_0)/\sqrt{1+\alpha^2(m-m_0)^2} ) $
 where $\beta$ is the maximum completeness, $m_0$ is the magnitude for $\beta/2$, 
and $\alpha$ represents the slope around the $m_0$, as shown by the solid lines in the figure.
The limiting magnitudes for 50\% photometric  completeness are F105W=30.44 mag and F220W=30.11 mag.

\section{Results}

\subsection{CMDs of the Galaxies}


In {\bf Figure \ref{fig:galaxyCMD}} 
we display the LW CMD (F277W vs (F277W--F356W)) and SW CMD (F200W vs (F150W--F200W)) (in AUTO magnitudes) of the extended sources in the central field of SMACS 0723. 
We also mark the cluster member galaxies (N=26) confirmed with spectroscopic observations \citep{mah22}, the selected bright galaxies (including the BCG), and the known gravitational lens image sources in \citet{mah22}.


In the F277W vs (F277W--F356W) CMD, two prominent  features are noted: an LW-blue vertical sequence with $(F277W-F356W)_0 \approx -0.6$,
and an LW-red slanted sequence with $(F277W-F356W)_0 \approx +0.1$. 
The known cluster members 
are mostly  located in the bright part of the LW-blue sequence. 
The known cluster members are mostly early-type galaxies, showing 
that this sequence corresponds to the  normal red sequence 
in the optical CMDs of galaxy clusters.
So these member galaxies were plotted in red circles.
The selected  bright galaxies including the BCG 
are also located in the bright part of this sequence.
The presence of a large number of galaxies located in this LW-blue sequence shows that SMACS 0723 is indeed a rich cluster.

In contrast, the known gravitational lens image sources are mostly located along the LW-red sequence, but with a larger scatter than the LW-blue sequence.  They look red in the released color images based on the combination of all SW and LW filter images. The gravitational lens image sources have much higher redshifts than that of SMACS 0723 so they are not the cluster members. They are background galaxies most of which show some star formation activity (i.e., late type galaxies having different colors compared with early-type galaxies).

However, the F200W vs (F150W--F200W) CMD
shows only one notable slanted sequence at (F150W--F200W)$_0\approx +0.2$. Both the cluster member galaxies and the lens image sources are located mostly in the same sequence, so they are hardly distinguished in this sequence.

Therefore, as the red sequence galaxies which are mostly early-type cluster members, we select those located along the LW-blue sequence as marked in the LW CMD:$-0.65 < (F277W - F356W)_0 < -0.45$.
We use them to check any correlation between the GCs and the galaxies in the following analysis. 

\subsection{CMDs of the Point Sources: Detection of the GCs}

{\bf Figure \ref{fig:pointCMD}} 
displays F200W vs (F150W--F200W) CMDs of the point sources in the central and  parallel fields. 
The CMD of the central field shows clearly a strong concentration of the point sources with GC colors  ($-0.2<(F150W-F200W)_0 <0.6$)
in the magnitude range of $26.5<F200W_0 < 30$ mag. 
  In contrast, a much smaller number of the sources are located in the same region of the CMD of the parallel field.
  This indicates that most of these sources in the central field belong to SMACS 0723.
  
  We plot also the GCs and other compact source candidates in the Image 2 field of 
Sparkler galaxy ($z=1.38$) located at $R=15\arcsec$ (82 kpc) east of the BCG in \citet{mow22}.
We cross-matched the compact sources in \citet{mow22} with our photometry and found that all the sources (N=12) were successfully recovered. However, only five of them passed our point source selection criteria. This indicates that a significant fraction of the compact sources in \citet{mow22} are at least marginally resolved or do not follow the clean JWST/NIRCam PSFs.
These sources are located in the similar GC color range, and the matched GCs occupy the brightest part of the GCs ($27<F200W_0 <28$ mag). 
Note that there are dozens more sources in the similar CMD region
($26.5<F200W_0 <28$ mag)
of the central field, but few are in the CMD of the parallel field. They are probably UCDs, but some of them may be candidates for
background GCs that were gravitationally lensed like the Sparkler galaxy GCs. 
The brightest of them are as bright as F200W$_0 \approx 26.5$ mag. 

In {\bf Figure \ref{fig:pointCMD}}  we also plot the color histograms of the point sources with 
$27.5 < F200W_0 <29.5$ 
mag in the central (thin red line) and parallel (thick black line) fields. Assuming the parallel field as the background, we subtract  the contribution of the background from the color histogram of the central field.  The resulting net color histogram of the central field (thick red line) shows a significant excess component that is close to Gaussian with a peak at $(F150W-F200W)_0 \approx 0.2$. 
This indicates that these point sources are mostly genuine GCs which belong to SMACS 0723. 

We select the GC candidate sample using the cyan box marked in the figure ($-0.2<(F150W-F200W)_0 <+0.6$ and $27.5<F200W_0 <29.5$ mag) for the following analysis. 
We conservatively selected the GC candidates with photometric completeness higher than ~70\% ($F200W_0 < 29.5$ mag).
The number of the selected GC candidates in the two fields is about 2500. 
{\bf Figure \ref{fig:finder}(c,d, and e)}
display  the zoomed-in images of the sample section of the BCG region that show these GCs better after subtracting the smooth background image from the original image.

The apparent magnitude range
($27.5<F200W_0 <29.5$ mag) of these sources at the cluster distance corresponds to the absolute magnitude range of 
$-14.2<M_{F200W}<-12.2$ mag, 
so these GCs belong to the bright end of the GCs, which include the UCDs.
Considering the redshift of SMACS 0723, the ages of these GCs are younger than 9.5 Gyr. 
The ages of most Milky Way GCs range from 11 Gyr to 13 Gyr, with a mean value of $12.3\pm0.4$ Gyrs \citep{oli20,kan21}). 
Thus  the SMACS 0723 GCs may be about 8 Gyrs old, assuming that they formed as the Milky Way GCs. 



We select two representative regions to check any difference between galaxy GCs and intracluster GCs: the BCG region (at $R<15\arcsec$ from the BCG center), and the intracluster GC region (with a radius of $15\arcsec$) that covers the western loop  \citep{mah22,pas22,mon22} and no bright galaxies, as marked in {\bf Figure \ref{fig:finder}(a)}. 
In {\bf Figure \ref{fig:GCCMD}(a,b)} 
we display F200W vs (F150W--F200W) CMDs 
of the point sources in the BCG region and the intracluster GC  region. 
We also plot the color histograms of both regions in {\bf Figure \ref{fig:GCCMD}(c)}. 
We derived the color histograms for the BCG and intracluster GC regions after subtracting the background contribution based on the parallel field. 
Note that the color distribution of the intracluster GC region shows a strong peak at $(F150W-F200W)_0 =0.05$ and a weaker component at  $(F150W-F200W)_0 \approx 0.4$.
In contrast the color distribution of the BCG region shows a strong peak at $(F150W-F200W)_0 =0.15$, which is  0.1 redder than the peak in the intracluster GC region. 

We calculate the blue GC fractions,  obtaining f(blue GC)=65\% for the intracluster GC region and 54\% for the BCG region. 
Thus the intracluster region has a higher fraction of blue GCs than the BCG region.

The color distribution for the intracluster GC region is broad across the peak color, showing that there are not only blue GCs but also red GCs, although the blue GC fraction is higher. The presence of both subpopulations in this region implies that the origin of the intracluster GCs is dual: the dominant progenitors of the intracluster GCs are low mass dwarf galaxies, and the minor progenitors are massive galaxies, as discussed in the case of Virgo intracluster GCs by \citet{lee10}.

Considering these, we divide the GCs in the entire field into two subpopulations according to their colors:
the blue (metal-poor) GCs with $-0.2<(F150W-F200W)_0 <+0.2$ and the red (metal-rich) GCs with $+0.2<(F150W-F200W)_0 <+0.6$.
We use these blue and red GCs  for the following analysis to study any difference between the two subpopulations.


\subsection{Spatial Distributions of the GCs}

{\bf Figure \ref{fig:GCmap}(a)} shows the spatial distribution of the GC candidates selected in the central field. 
In  {\bf Figure \ref{fig:GCmap}(b)} we plot the number density contours and color gradient map of these sources on the gray scale map of the F200W image. The contours were smoothed with a Gaussian smoothing scale, $\sigma_G \simeq3\arcsec$.
The lowest level in the contours starts with  $2\sigma_{bg}$ where $\sigma_{bg}$ is estimated from the background fluctuation of the selected GC candidates in the low density region of the parallel field: $\sigma_{bg} = 0.00052 \pm 0.000029$ per kpc$^2$.
The step between contours is  $4\sigma_{bg}$.
We also mark the locations of the known cluster members, the selected brightest galaxies (including the BCG), the gravitational lens image sources \citep{mah22}, and the red sequence galaxies 
selected in this study, 
to check any spatial correlation with the GCs. 

Several features are noted in this figure.
First, the overall distribution of the GC candidates shows clearly a central concentration around  the BCG that is located close to the galaxy cluster center. This implies that most of the selected GC candidates are the members of SMACS 0723, indeed.
The strongest peak of the GC number density is slightly offset to the east direction from the BCG center, and this can be explained as follows.
There are two galaxies close to the BCG center in the west direction so it is difficult to subtract completely the stellar  light of these galaxies in the central region. This decreases the detection rate of the GCs in that region, causing the slight offset from the BCG center. However the center of the contours at $R>10\arcsec$ is consistent with the BCG center.

Second, the overall distribution of the GCs is elongated along the major axis of the BCG.
Third, the spatial distribution of the GCs is much more extended than that of the stellar light of the BCG. 

Fourth, strong peaks are found at the position of some brightest galaxies. The second strongest peak of the GC number density is seen at the position of G2, a bright elliptical member galaxy 
($z=0.3841$) 
at $R\approx 190$ kpc in the northwest of the BCG.
Another weaker peak is seen at the position of G3, a bright galaxy close to G2. 
A recognizable clumping of GCs is seen also at the position of G4, another elliptical member galaxy ($z=0.3845$). 
There is an isolated weak GC clump at $R\approx 300$ kpc south of the BCG, at the center of which one bright red sequence galaxy is located.
The rest of the known cluster member galaxies do not show any strong clumps of GCs.

Fifth, there is a strong concentration of GCs at $R\approx 200$ kpc in the east of the BCG (called as the GC Clump 1 (GCC 1)). Interestingly there are no recognizable bright galaxies which can be a host of these GCs. 
Similarly there is a weak concentration of GCs at $R\approx 100$ kpc in the west of the BCG (called as the GC Clump 2 (GCC 2)). There are no recognizable bright galaxies which can be a host of these GCs. 

Sixth, there is seen a concentration of GCs extended to the southwest from the second strongest G2 peak. 
This can be a part of the large western loop-like structure, which will be described in detail later. 
We selected the intracluster GC region to cover this substructure, as marked by the large circle in the figure. 

Finally, the GCs located outside the few bright red sequence galaxies (including the BCG) are considered to be mostly intracluster GCs. 

In {\bf Figure \ref{fig:GCmapBR}} we display the spatial distributions of the blue GC and red GC candidates selected in the central field. 
This figure shows that the central concentration of the red GCs around the BCG is more prominent than that of the blue GCs, which will be also shown by their radial number density profiles in the following.
In addition, these maps show a loop-like structure with a radius of 100 kpc at $R\approx 200$ kpc west of the BCG. The structure is seen better in the red GC map.
This will be compared with the substructure in the ICL map in the following.

\subsection{Radial Number Density Profiles of the GCs}

Adopting the BCG center as the cluster center we derive the radial number density profile of the GCs, displaying it in {\bf Figure \ref{fig:GCRDP}(a)}.
The number density profiles of the GCs derived before background subtraction show a flattening at  $R>90\arcsec$  in the parallel field.
We estimate the background level from the sources at $R>90\arcsec$ in the parallel field,  
obtaining $\Sigma_{bg}= 0.000831\pm0.000075$ per kpc$^2$. 
In the figure the number density of the GCs deceases as the cluster-centric distance increases, confirming again that these GCs mostly  belong to the galaxy cluster. The radial number density of the GCs keeps decreasing until $R\approx 600$ kpc.
The central excess component represents the BCG GCs, while the outer part is dominated by the intracluster GCs.

In {\bf Figure \ref{fig:GCRDP}(b)} we also plot the radial number density profiles of the blue and red GCs. 
The radial profile of the red GCs shows a stronger excess in the central region than the blue GCs,
and it shows an opposite trend in the outer region.

We fit the radial number density profiles of the GCs 
with the two component S\'ersic law (the BCG component plus the intracluster GC component): $\Sigma_{GC}(R) = \Sigma_{\rm eff} \exp{ \{-b_{n}[(R/R_{\rm eff})^{1/n} - 1]\} }$ 
and    $(b_{n} = 2n - 0.3271)$ \citep{ser63,gra05}.
We also fit the data with the power law for  the outer region of the intracluster GC component ($158<R<631$): 
$\Sigma_{GC}(R)\propto R^\alpha$.
Table \ref{tab:fitRDP} lists a summary of the fitting results.
The radial profile of all GCs is fit well with the S\'ersic parameters, 
$R_{\rm eff,1}=28.5\pm3.2$ pc and $n_1=0.1\pm0.1$ for the BCG component, and $R_{\rm eff,2}=224.5\pm22.1$ pc, $n_2=1.3\pm0.3$ for the intracluster component. 
The outer region 
is also fit well with a power law index  $\alpha=-2.0\pm0.2$. 
The radial profiles of the blue GCs and red GCs are also fit with similar parameters:
in the case of the blue GC,  $R_{\rm eff,1}=29.5\pm6.5$ pc, $n_1=0.1\pm0.2$ for the BCG component, and  $R_{\rm eff,2}=213.7\pm22.1$ pc, $n_2=1.1\pm0.3$ for the intracluster GC component, and   $\alpha=-2.2\pm0.2$ for the outer region;
in the case of the red GCs, $R_{\rm eff,1}=26.7\pm3.4$ pc, $n_1=0.1\pm0.1$ for the BCG component $R_{\rm eff,2}=241.7\pm48.6$ pc, $n_2=1.5\pm0.6$ for the intracluster GC component, and  $\alpha=-1.9\pm0.3$ for the outer region.

Thus the effective radius of the intracluster GC system is about eight times larger than that of the BCG GC system. 
The outermost region of the intracluster GC system 
is also represented well by the power law with an index of $\alpha = -2.0\pm0.2$. This profile is slightly steeper than that of the intracluster GC system in the Virgo cluster, $\alpha= -1.5\pm0.1$\citep{lee10}.





\section{Discussion}

\subsection{Comparison with ICL Distribution}

 From the JWST/NIRCam images of SMACS 0723, \citet{pas22} and \citet{mah22} derived a smooth map of the diffuse light using a median box car filter with $21 \times 21$ pixels and $100 \times 100$ pixels, respectively, showing the distribution of the ICL.  
 They noted that the diffuse ICL is elongated along the major axis of the BCG, but in a much more extended way. Also they found two interesting substructures of the ICL: a large scale loop in the west of the BCG and a large lobe-like feature in the east component.
 
 To compare the distributions of the GCs and ICL, we derive similarly a smooth map of the diffuse light applying a median boxcar smoothing with $51 \times 51$ pixels to the F277W image,
 and display it in 
 {\bf Figure \ref{fig:ICLmap}(a)}. 
 We overlay the GC number density contour maps (all GCs, blue GCs, and red GCs) to the smooth map of the ICL
 in {\bf Figure \ref{fig:ICLmap}(b,c, and d)}.
 We also mark the positions of 
 bright stars, known cluster member galaxies, and gravitational lens image sources \citep{mah22} in the figure.
 
{\bf Figure \ref{fig:ICLmap}} shows several distinguishable features.
First, the spatial distributions of the GCs and ICL are consistent in general.
However, the GCs are found farther than the ICL boundary.
Second, the western large scale loop structure seen in the GC number density map
is consistent with that of the ICL. A number of GCs are located along the loop of the diffuse ICL. %
Third, the GC number density map shows also an extended structure in the east of the BCG, which is consistent with the lobe-like diffuse component of the ICL. However, the GC Clump 1 (the strongest clump of GCs at $R\approx 200$ kpc east of the BCG)  does not have a bright counterpart in the ICL map, although it is close to the boundary of the east loop.
The GC Clump 1 is better seen in the blue GC map, indicating that it is composed mainly of the blue GCs.

\subsection{Formation of SMACS 0723}

The existence of a number of substructures with varying scales (in addition to the BCG) in the GC map and ICL map indicates that a large number of galaxy groups with diverse masses are falling to the cluster center, and they are in the middle of merging process now. 
The elongated structure seen in the maps of the GCs  and ICL indicates that most galaxies are falling to the cluster center along the east-west direction.
These maps illustrate the distribution of the stars and GCs that are stripped off from their host galaxies or are the remnant of disrupted galaxies.

Considering the distributions of the GCs and ICL, G2 is probably a main host of a galaxy group that includes G3 (called the G2 group). 
{\bf Figure \ref{fig:ICLmap}}  shows that the G2 group and the BCG are connected not only in the GC map but also in the ICL map. The existence of the GCs and diffuse light between the G2 group and the BCG implies that both systems are physically interacting. This is similar to the case of the NGC 4839 group that is known to be falling to the Coma center, as shown in the observations of galaxies and various ICM including the GCs
\citep{bri92, whi93,col96,lys19,chu21,oh22}.
The loop structure might have formed when the G2 group is falling to the cluster center, which needs to be investigated further in the future.

\subsection{Comparison with X-ray Brightness Distribution}

Chandra X-ray brightness map of SMACS 0723 is given in \citet{mah22} (their Figs. 1 and 9).
Comparing the GC number density map in this study with the X-ray map of SMACS 0723 given in \citet{mah22},
we find both maps display generally similar structures in the sense that both  show a large scale elongated distribution.
Interestingly the X-ray contour map of the central region of SMACS 0723 (Fig. 1 in \citet{mah22}) shows
 a distinguishable component at the position of the western loop structure in addition to the strongest peak at the BCG center.
The location of this component is consistent with the center of the western loop in the GC number density map and ICL map.

\subsection{Comparison with Dark Matter Distribution}

Dark matter mass density maps (and radial profiles) of SMACS 0723 derived from the strong lensing analysis were presented by \citet{mah22,pas22,cam22, gol22} who used the JWST/NIRCam images 
and HST/ACS/WFC3 images.
 In {\bf Figure \ref{fig:compmassmap}} we overlay  the GC number density contour maps on the dark matter surface mass density map (the convergence ($\kappa$=surface mass density normalized to the critical mass density) map) provided by \citep{mah22} (pseudo color maps).  
The GC maps and dark matter mass maps show remarkably similar structures in the sense that both  show a large scale structure that is elongated along the major axis of the BCG, but much more extended than the BCG. They show also similar substructures. 

\citet{mah22} present a comparison of the dark matter radial density profiles of SMACS 0723 they derived from the JWST/NIRCam images with those derived using different lensing models in the literature:
RELICS-lenstool \citep{coe19}, RELICS-GLAFIC \citep{coe19}, LTM \citep{gol22} (their Fig. 6). 
In this comparison, the first three models produce similar profiles extending out to $R\approx 1$ Mpc, while the LTM profile is much more steeply declining in the outer region ($R\approx 500$ kpc).

In {\bf Figure \ref{fig:compRDP}} we present a comparison of the radial number density profiles of the GCs with the dark matter mass density profiles based on the lensing analysis \citep{mah22}. 
This figure shows that the radial density profile of the GCs is very similar to those of the dark matter in the inner range of $R=50-200$ kpc. However it becomes steeper in the outer region at $R=200-600$ kpc,  deviating from the three lens models (RELICS-lenstool, RELICS-GLAFIC, and \citet{mah22}) and getting closer to the LTM model. However, the outermost GC profile  is between the two model groups.

These results are consistent with those for the GCs found in the deep HST images of Abell 2744, another massive cluster at $z=0.308$ that is known as one of the most GC-abundant galaxy clusters \citep{lee16a}.  
Similar trends are also seen in the radial profiles of the intracluster GC number density and the dark matter mass derived from the 
simulated galaxy clusters selected from the cosmological simulations \citep{ram18, ram20}.



\section{Summary}

In the search of the GCs in the massive galaxy cluster SMACS 0723 using the early released  JWST/NIRCam images, we find a large population of GCs which  spread over the cluster. These GCs are mostly intracluster GCs as well as the BCG GCs. 
The ages of these GCs, considering the cluster redshift, are younger than 9.5 Gyr. These GCs are about 8 Gyr old, if they formed as the Milky Way GCs.

Primary results are summarized as follows.
\begin{enumerate}
\item The F200W magnitudes of these GCs, $26.5<F200W_0 <29.5$ mag, correspond to $-15.2<M_{F200W}<-12.2$ mag, showing that they belong to the brightest GCs (including the UCDs)). 

\item The spatial distributions of these GCs  show a megaparsec-scale structure that is elongated along the major axis of the brightest cluster galaxy, but is much extended than the stellar light of the BCG.  
In addition, they show a large number of substructures with various scales,
some of which are consistent with the substructures in the map of the diffuse ICL. 

\item The existence of the GCs and ICL between the G2 group and the BCG indicates that the G2 group is falling to the cluster center, as in the case of the NGC 4839 group in the Coma cluster. 
The large scale loop structure in the west of the BCG might have formed when the G2 group is falling to the cluster center.

\item The GC number density map is, in general, consistent with that of the dark matter mass density map based on the strong lensing analysis in the literature in the sense both show a large-scale elongated structure.

\item The radial number density profiles of the GCs 
are fit reasonably by the two component S\'ersic law: the BCG component and the intracluster GC component. The effective radius of the intracluster GC system is about eight times larger than that of the BCG GC system. 

\item The radial number density profile of the GCs
in the inner region at $R=50-200$ kpc is very similar to the dark matter mass profile obtained from strong lensing models. 
However it becomes steeper in the outer region at $R=200-600$ kpc,  deviating from the three lens models 
and getting closer to the LTM model. However, the outermost GC profile is between the two model groups.

\end{enumerate}

All these results, based on the excellent data obtained with the JWST/NIRCam, show that the intracluster GCs are an excellent independent tool to probe the dark matter distribution in galaxy clusters as well as to reveal the cluster assembly history. SMACS 0723 is only the first case of distant galaxy clusters showing the power of the JWST, and many more will come in the JWST era.

\acknowledgments
The authors are grateful to the anonymous referee for useful comments. This work  was supported by the National Research Foundation grant funded by the Korean Government (NRF-2019R1A2C2084019). 
We thank Brian S. Cho for improving the English in the original manuscript.
This work is based on observations made with the
NASA/ESA/CSA James Webb Space Telescope.
The data
were obtained from the Mikulski Archive for Space Telescopes
at the Space Telescope Science Institute, which is
operated by the Association of Universities for Research
in Astronomy, Inc., under NASA contract NAS 5-03127
for JWST. 

\vspace{15mm}
\facilities{JWST}

\software{ Numpy \citep{har20b}, Matplotlib \citep{hun07}, Scipy \citep{vir20}, Astropy \citep{ast13, ast18}, IRAF\citep{tod86}, PyRAF \citep{sts12}, 
Photutils \citep{bra21}, Astroalign \citep{ber20}, DAOPHOT \citep{ste87}, and SExtractor \citep{ber96} }

Some/all of the data presented in this paper were obtained from the Mikulski Archive for Space Telescopes (MAST) at the Space Telescope Science Institute. The specific observations analyzed can be accessed via \dataset[DOI]{https://doi.org/DOI}.



\clearpage

\begin{deluxetable}{ccccccccc}
\tablecolumns{11}
\tablewidth{0pc}
\tablecaption{Basic Properties of SMACS 0723} 
\label{tab:info}
\tablehead{
\colhead{Parameter} & \colhead{Value} & \colhead{References}  }
\startdata
Heliocentric Velocity & 116919 \kms & 1,2 \\ 
Redshift & $z=0.39$ & 1,2 \\ 
Luminosity Distance$^a$  & $2170$ Mpc & 1 \\
Distance Modulus$^a$ & $(m-M)_0  = 41.69\pm0.15$ & 1 \\ 
Image Scale$^a$ & 5.453 kpc arcsec$^{-1}$ = 327.16 kpc arcmin$^{-1}$ & 1 \\ 
Age at Redshift$^a$  &  9.474 Gyr & 1 \\ 
Foreground Extinction &  $A_V=0.585$, $A_I=0.328$ & 3 \\ 
& $A_{F090W}=0.296$, $A_{F150W}=0.134$, $A_{F200W}=0.083$ & 3 \\
& $A_{F277W}=0.053$, $A_{F356W}=0.039$, $A_{F444W}=0.031$ & 3 \\ 
Velocity Dispersion &$\sigma_v = 1,180\pm170$ \kms & 2 \\ 
Mass  & $M(R<400 {\rm kpc})=3\times 10^{14} M_\odot$& 2 (Lensing) \\ 
 & $M(R<1 {\rm Mpc})=8\times 10^{14} M_\odot$ & 4 (PLANCK) \\ 
\enddata
\label{tab_a2744basic}
\tablenotetext{a}{NED values for $H_0 = 67.8$ \kmsMpc, $\Omega_M=0.308$, $\Omega_\Lambda = 0.692$.}
\tablecomments{References: (1) the NASA/IPAC Extragalactic Database (NED); (2) \citet{mah22};
(3) \citet{sch98,sch11}; (4) \citet{pas22}.}
\end{deluxetable}
\clearpage
%
\begin{deluxetable}{cccc}
\tablecolumns{11}
\tablewidth{0pc}
\tablecaption{Fitting Results of the GC Radial Number Density Profiles} 
\label{tab:fitRDP}
\tablehead{
\colhead{GC sample} & \colhead{BCG component$^a$} & \colhead{Intracluster GC component$^a$} & \colhead{Power Law Slope$^b$}  }
\startdata
All GC & $R_{\rm eff,1}=28.5\pm3.2$ pc, $n_1=0.1\pm0.1$ & $R_{\rm eff,2}=224.5\pm22.1$ pc, $n_2=1.3\pm0.3$ & $\alpha=-2.0\pm0.2$\\
Blue GC & $R_{\rm eff,1}=29.3\pm7.0$ pc, $n_1=0.1\pm0.2$ & $R_{\rm eff,2}=238.9\pm29.1$ pc, $n_2=1.3\pm0.3$ & $\alpha=-2.1\pm0.2$ \\
RGC GC & $R_{\rm eff,1}=26.7\pm3.4$ pc, $n_1=0.1\pm0.1$ & $R_{\rm eff,2}=241.7\pm48.6$ pc, $n_2=1.5\pm0.6$ & $\alpha=-1.9\pm0.3$ \\
\enddata
\tablenotetext{a}{Fitting with the two component S\'ersic law (the BCG component plus the intracluster GC component): 
$\Sigma_{GC}(R) = \Sigma_{\rm eff} \exp{ \{-b_{n}[(R/R_{\rm eff})^{1/n} - 1] \}} $  and    $b_{n} = 2n - 0.3271$  \citep{ser63,gra05}.}
\tablenotetext{b}{Fitting with the power law for  the outer region ($158<R<631$ kpc) of the intracluster GC component: $\Sigma_{GC}(R)\propto R^\alpha$.}
\end{deluxetable}
\clearpage

\begin{figure*}
    \centering
    \plottwo{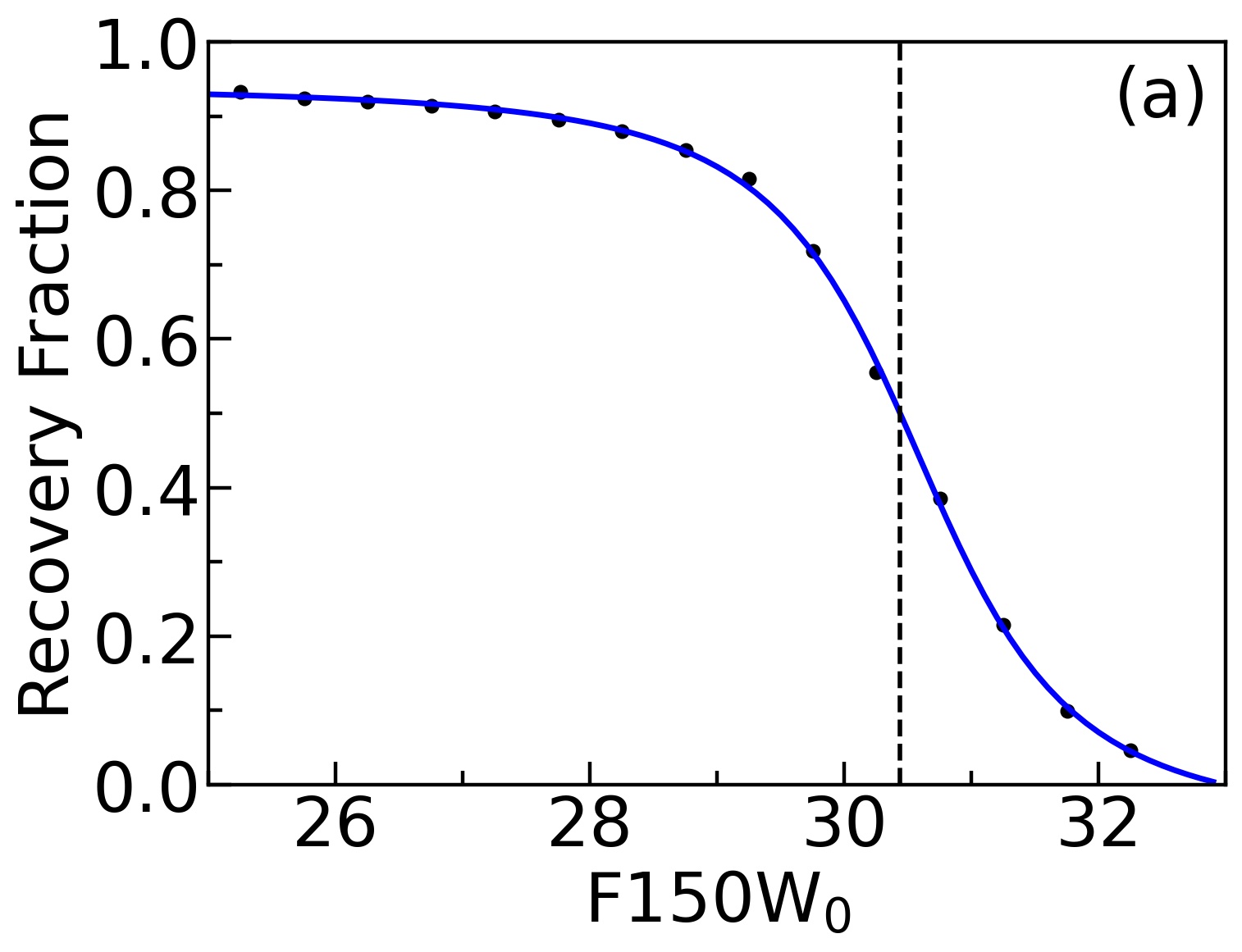}{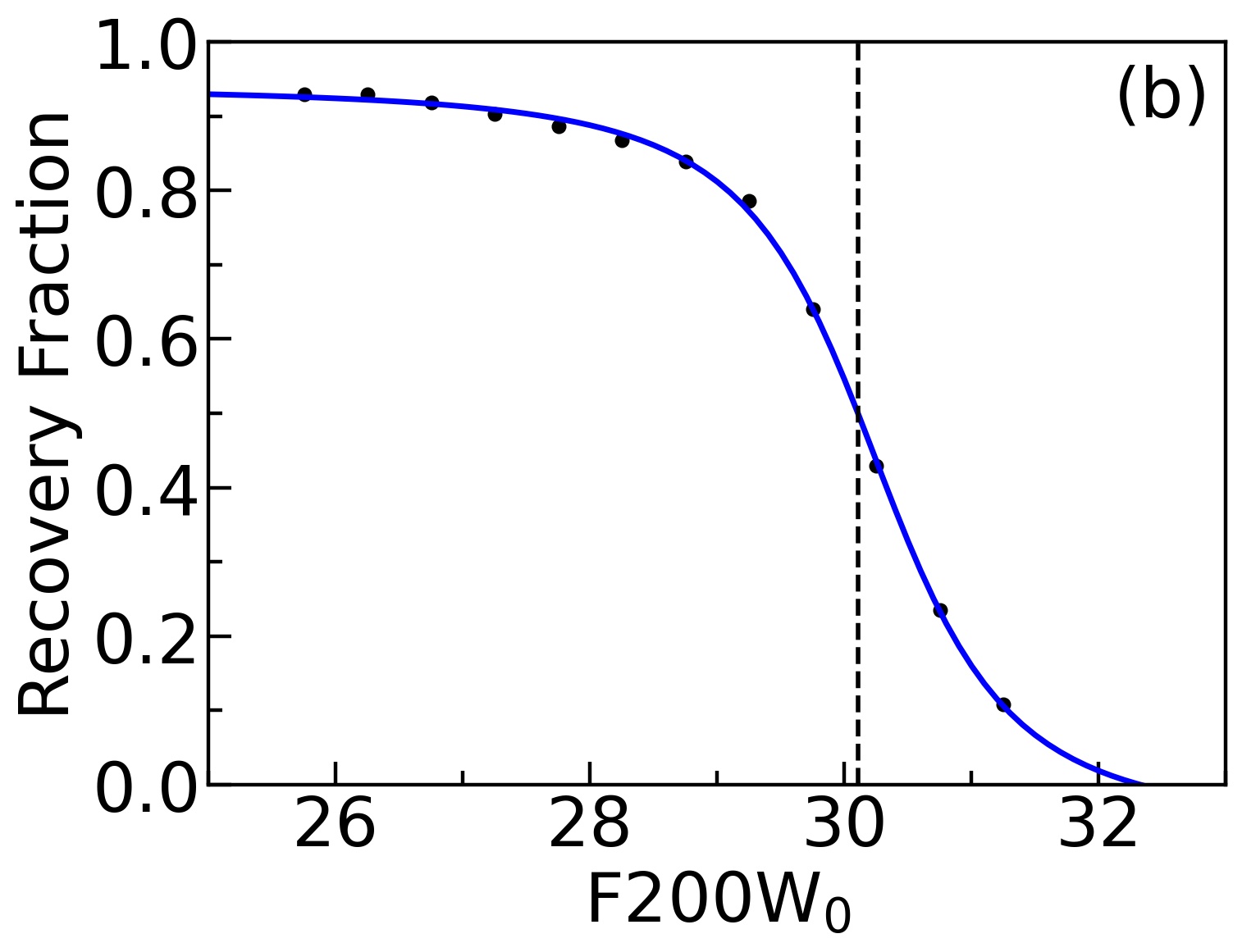}
    \caption{
    Completeness (recovery fraction) vs. F150W (a) and F200W (b) magnitudes of the point sources. Blue solid lines denote the fitting results.
    The vertical black lines mark 50\% completeness limits: $F150W_{\rm lim}= 30.44$ mag, and $F200W_{\rm lim}= 30.11$ mag.
     }
	\label{fig:completeness}
\end{figure*}
\clearpage

\begin{figure*}
    \centering
    \plottwo{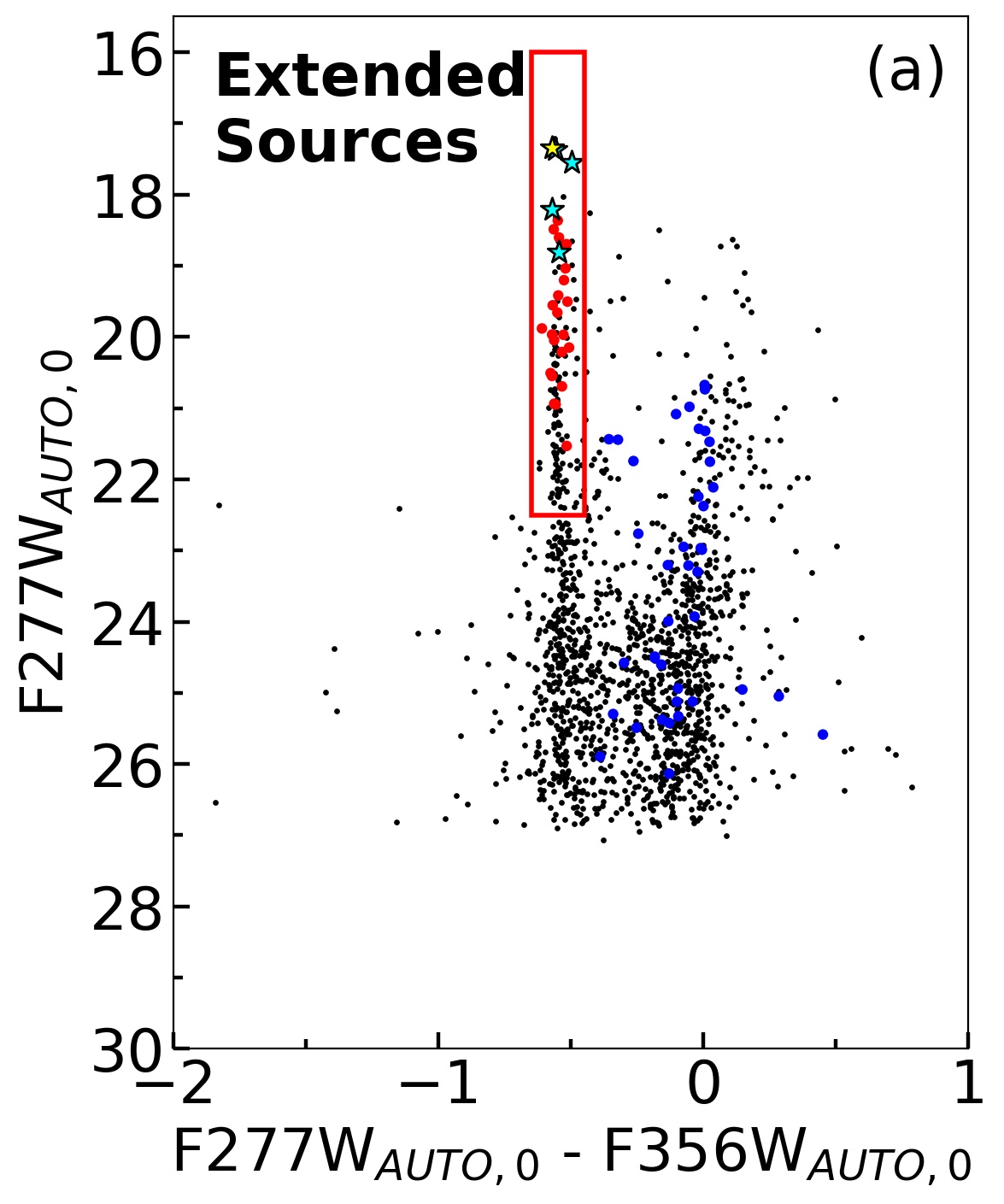}{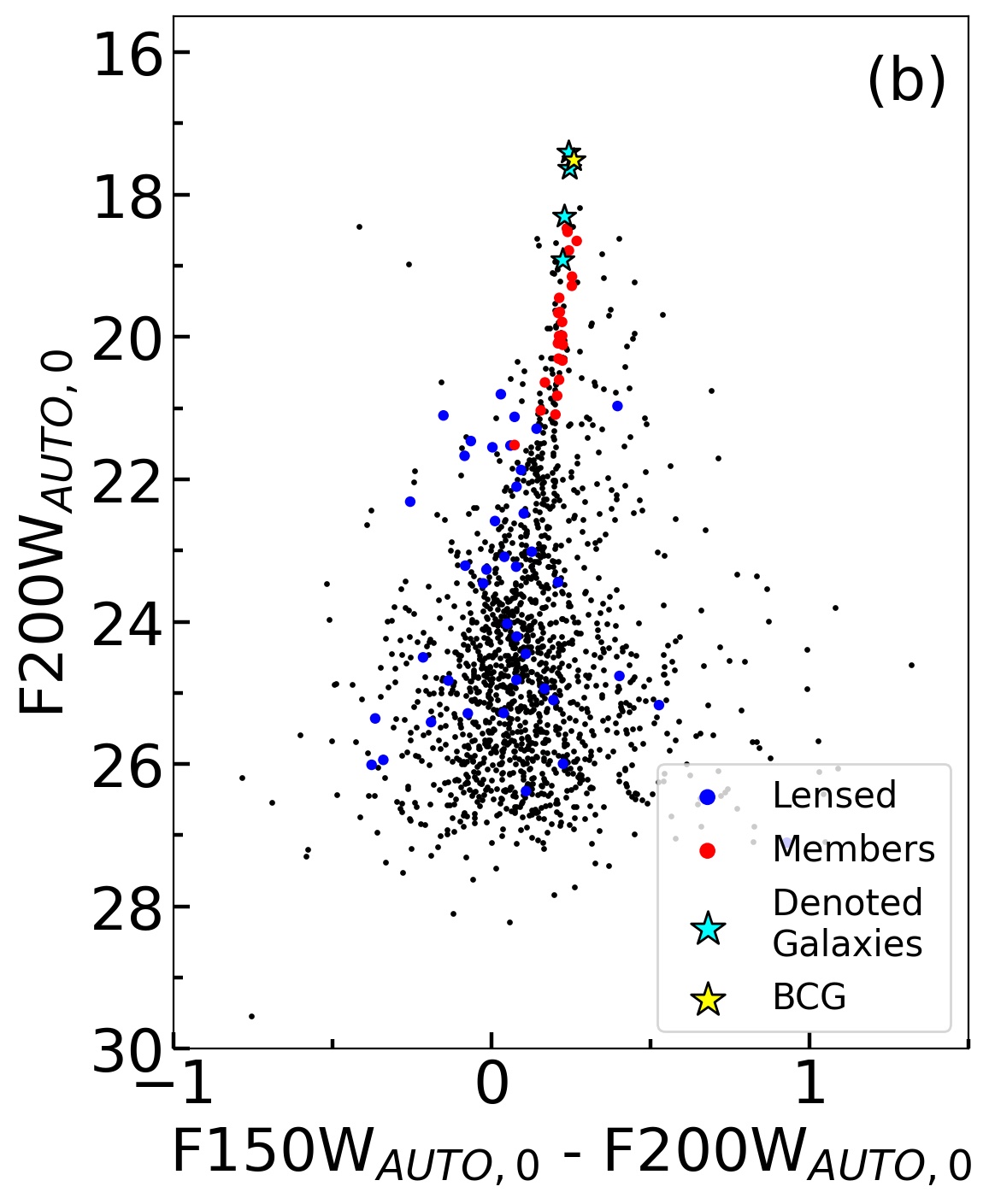} 
    \caption{LW CMD (F277W vs. (F277W--F356W)) and SW CMD (F200W vs. (F150W--F200W) of the extended sources 
    with AUTO magnitudes  in the central field
    of SMACS 0723.
    Red circles denote the spectroscopically confirmed cluster member galaxies, and blue circles represent the gravitational lens image sources \citep{mah22}.
    The yellow starlet is the BCG, and cyan starlets represent the bright galaxies (G2, G3, G4, and G5) marked in Figure \ref{fig:finder}. 
        The red box in the left CMD represents the selection criteria for the optically-red sequence galaxies which are mostly early-type members of the cluster.
        Note that the blue member galaxies in (F277W--F356W) colors are optically red galaxies so they are plotted with red circles.}  
	\label{fig:galaxyCMD}
\end{figure*}
\clearpage

\begin{figure*}
    \centering
    \includegraphics[scale=0.17]{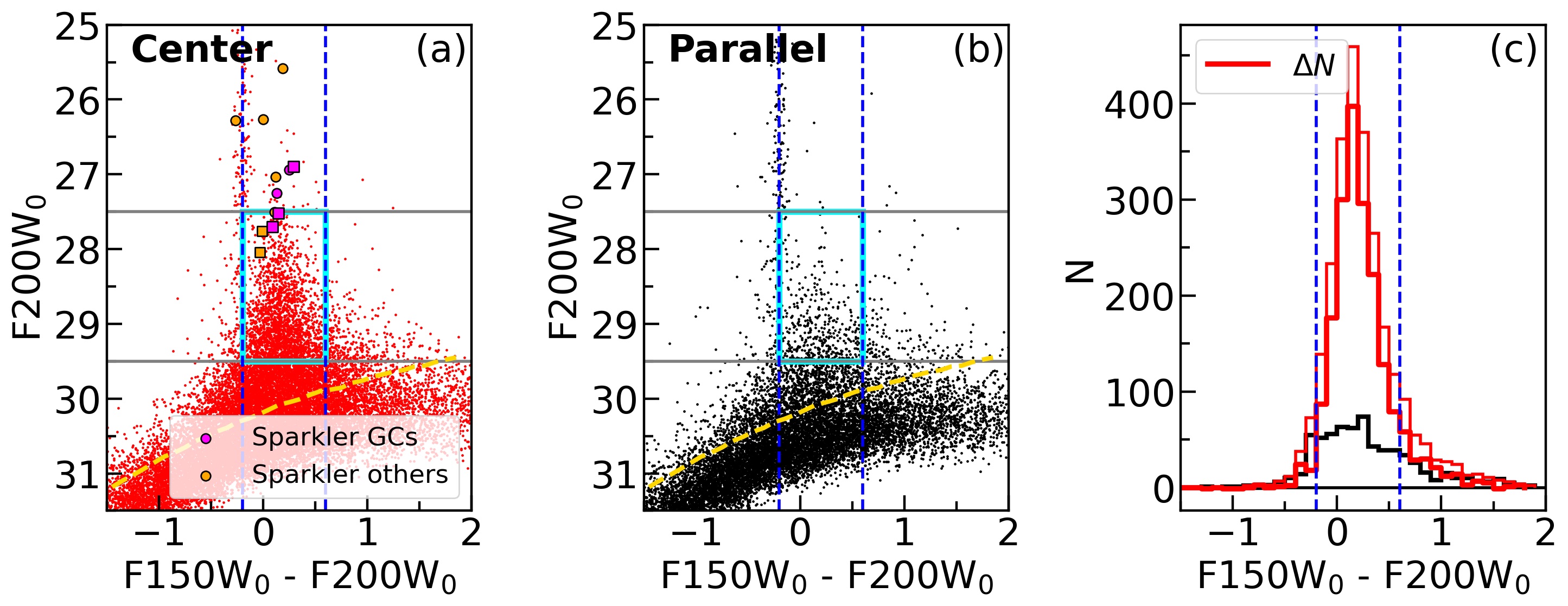} 
    \caption{
   F200W vs. (F150W--F200W) CMDs of the point sources in the central (a) and parallel (b) fields of SMACS 0723.
    Magenta and orange circles, respectively, represent the GCs and other compact sources of the Sparkler galaxy ($z=1.38)$ in \citet{mow22}.
    The sources inside the cyan box of the central field are mostly GCs belonging to SMACS 0723.
    Cyan boxes represent the boundary for selecting the GC candidates in this study.
   The curved yellow dashed lines mark the 50\% completeness limit.
     In (c) the thick red line denotes the net color histogram of  the point sources with $27.5<F200_0<29.5$ mag in the central field (thin red line) after subtracting the contribution of the background based on the parallel field (black line). 
    }
	\label{fig:pointCMD}
\end{figure*}

\begin{figure*} 
    \centering
    \includegraphics[scale=0.17]{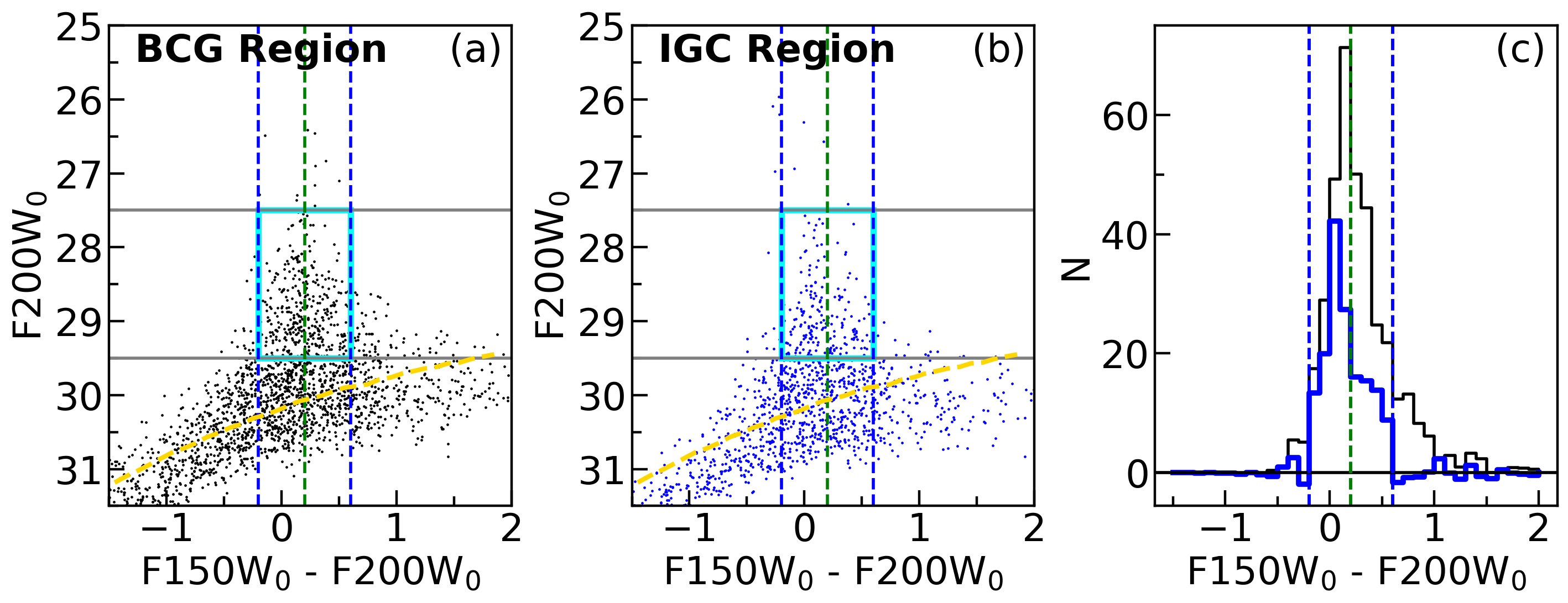} 
    \caption{(a,b) F200W vs. (F150W--F200W) CMDs  of the point sources in the BCG region 
    and the intracluster GC region 
    in the central field of SMACS 0723, as marked in Figure  \ref{fig:finder}. 
    The curved yellow dashed lines mark the 50\% completeness limit.
    (c) Color histograms of the BCG region (black line) and the intracluster GC region (blue line) after subtracting the contribution of the background based on the parallel field. 
   Note that the color distribution of the intracluster GC region shows a strong peak at $(F150W-F200W)_0 =0.05$, while that of the BCG region shows a strong peak at 0.1 redder color,  $(F150W-F200W)_0 =0.15$. Note also both regions show a broad color distribution across the peak.
    Cyan boxes represent the boundary for selecting GC candidates, and green dashed lines mark the boundary for separating blue and red GCs.
    }
	\label{fig:GCCMD}
\end{figure*}
\clearpage

\begin{figure*}
    \centering  
    
 \includegraphics[scale=0.17]{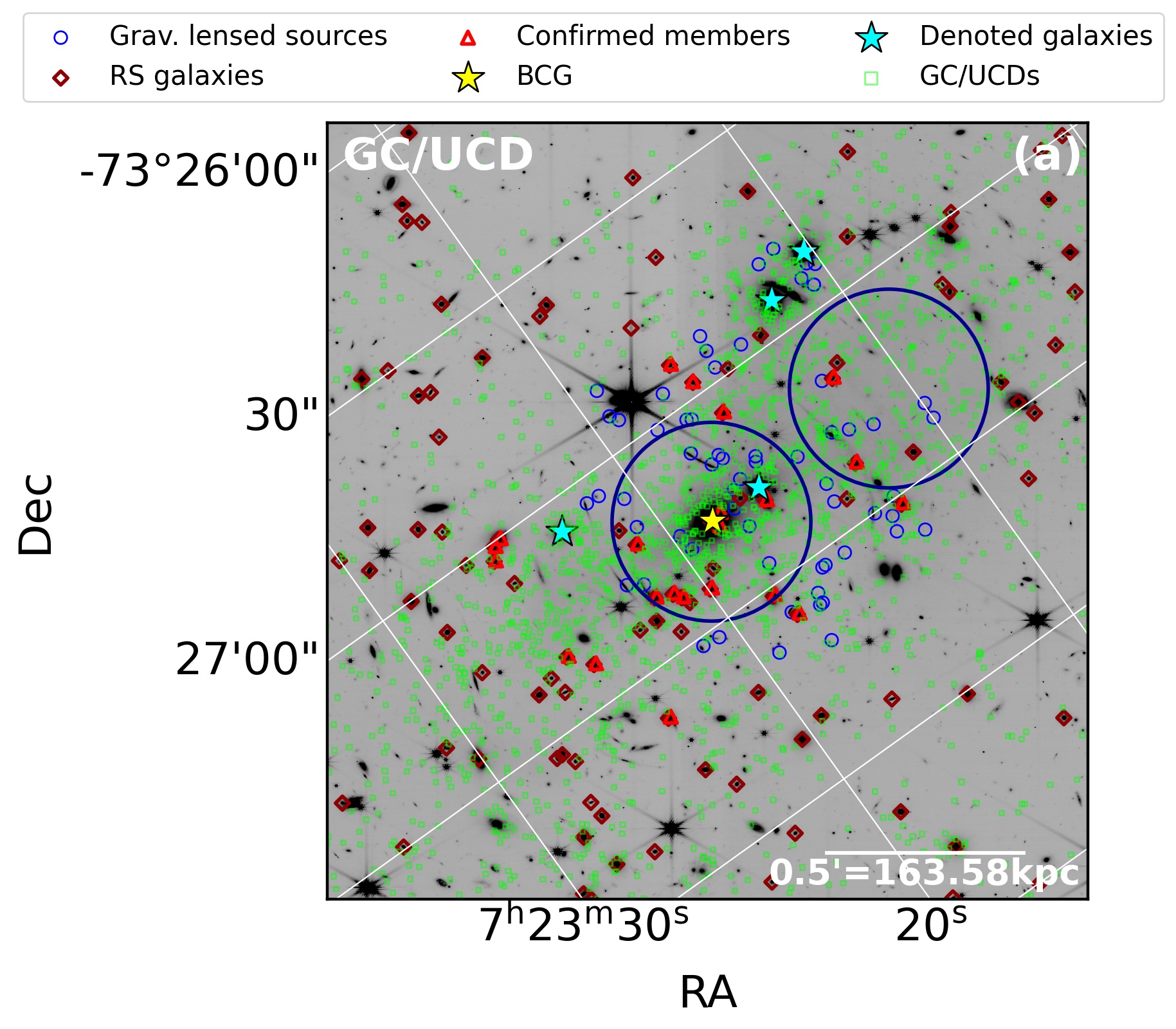} 
\includegraphics[scale=0.17]{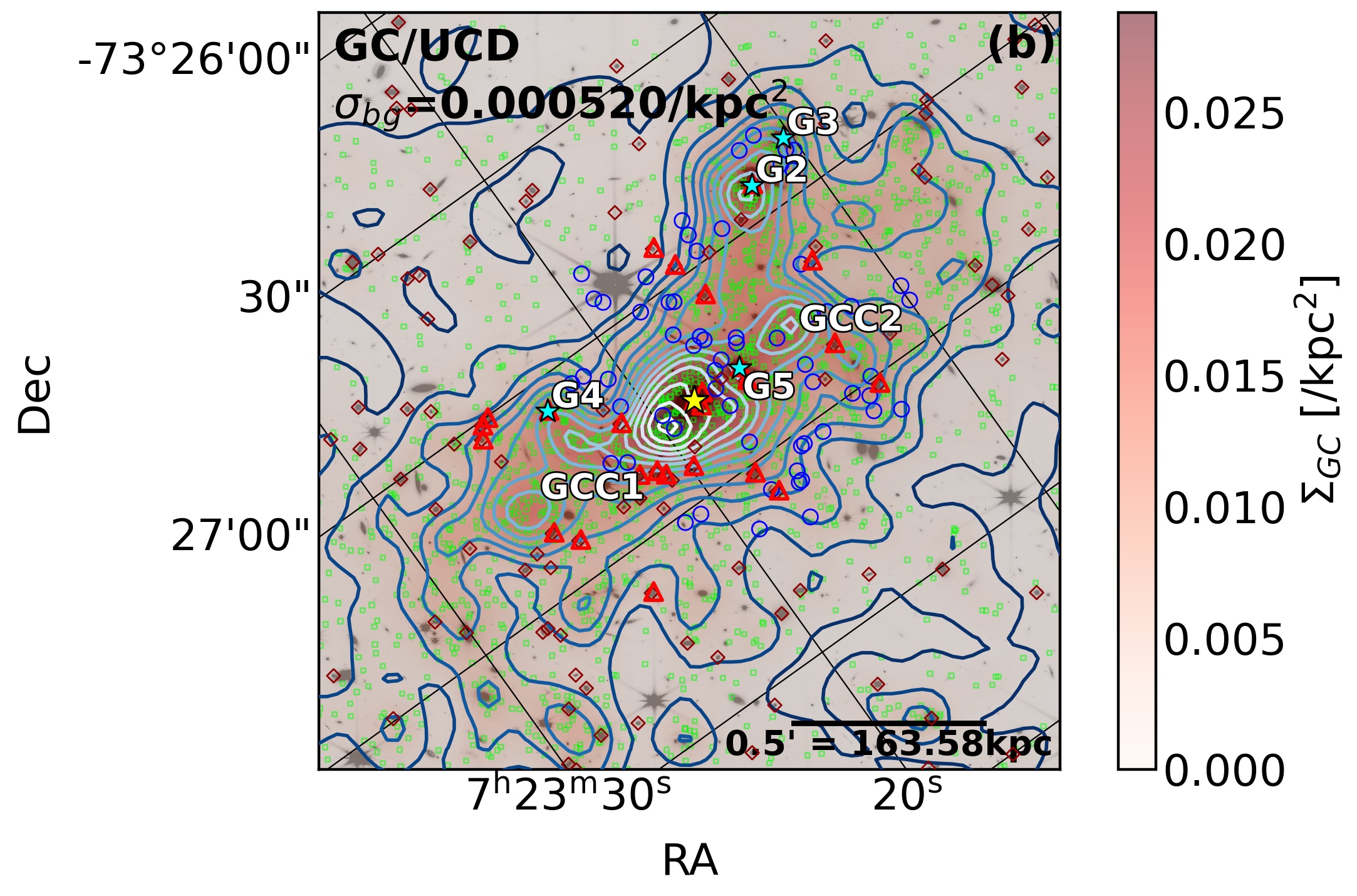} 
    \caption{(a) Spatial distribution  of the GCs (green squares)  in the gray scale map of the F200W image for the central field of SMACS 0723.
    Red triangles represent
    the confirmed cluster member galaxies, and  dark red diamonds denote the selected bright red sequence galaxies (with $-0.65 < (F277W - F356W)_0 < -0.45$ and F277W$_0< 22.5$ mag).
    Blue open circles represent the gravitational lens image sources \citep{mah22}. 
    Yellow and cyan starlets denote the BCG, and other bright galaxies.
    Two large circles mark the BCG region and the intracluster GC region.
    (b) The contour map and color gradient map of the GC number density in the gray scale map of the F200W image.  
    The contours were plotted from 2$\sigma_{bg}$ level with $4\sigma_{bg}$ step.
    The color scale bar represents the GC number density per kpc$^2$.
    GCC1 and GCC2 denote GC clumps.}
	\label{fig:GCmap}
\end{figure*}

\begin{figure*}
    \centering  
\includegraphics[scale=0.2]{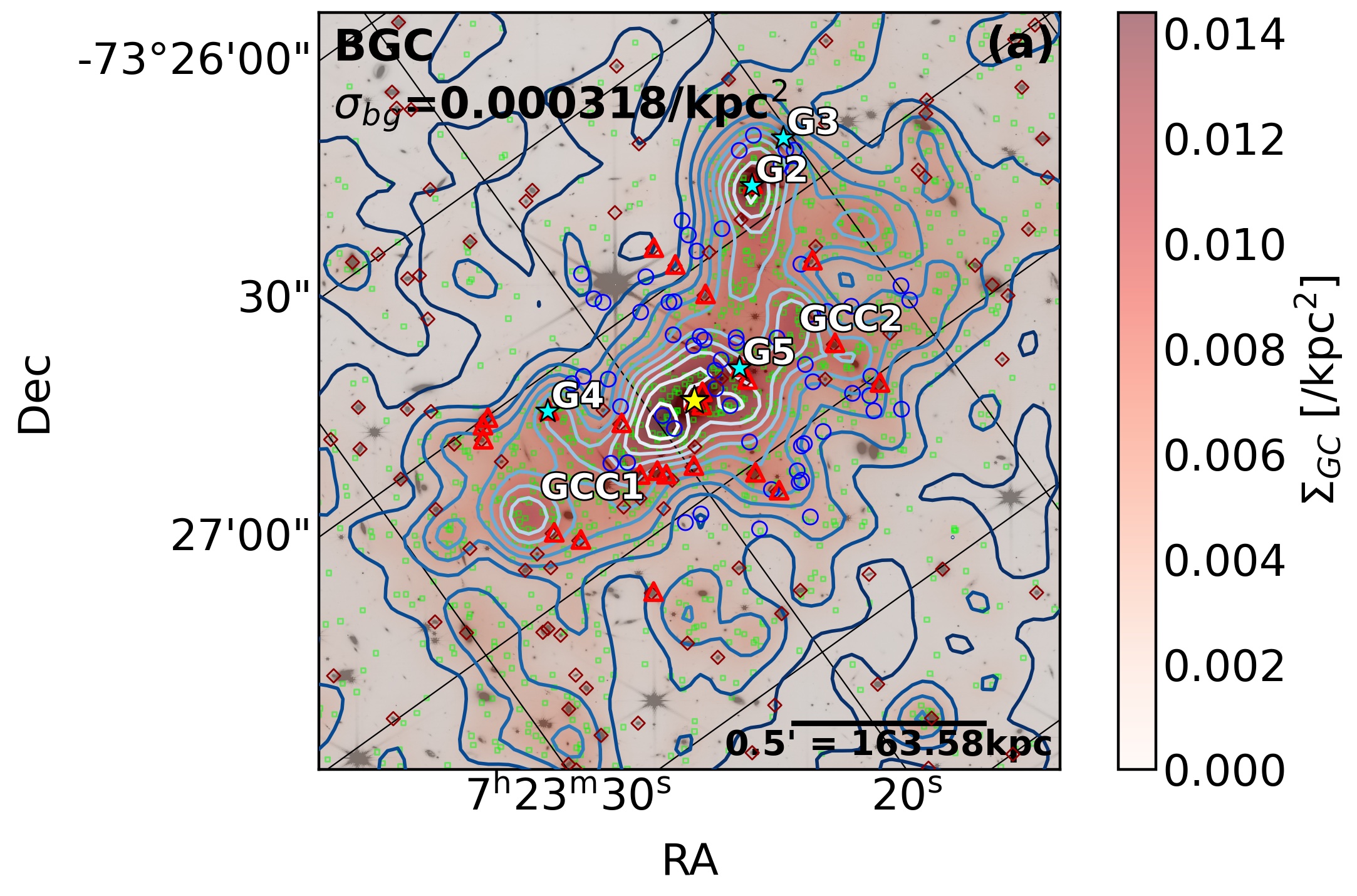} 
\includegraphics[scale=0.2]{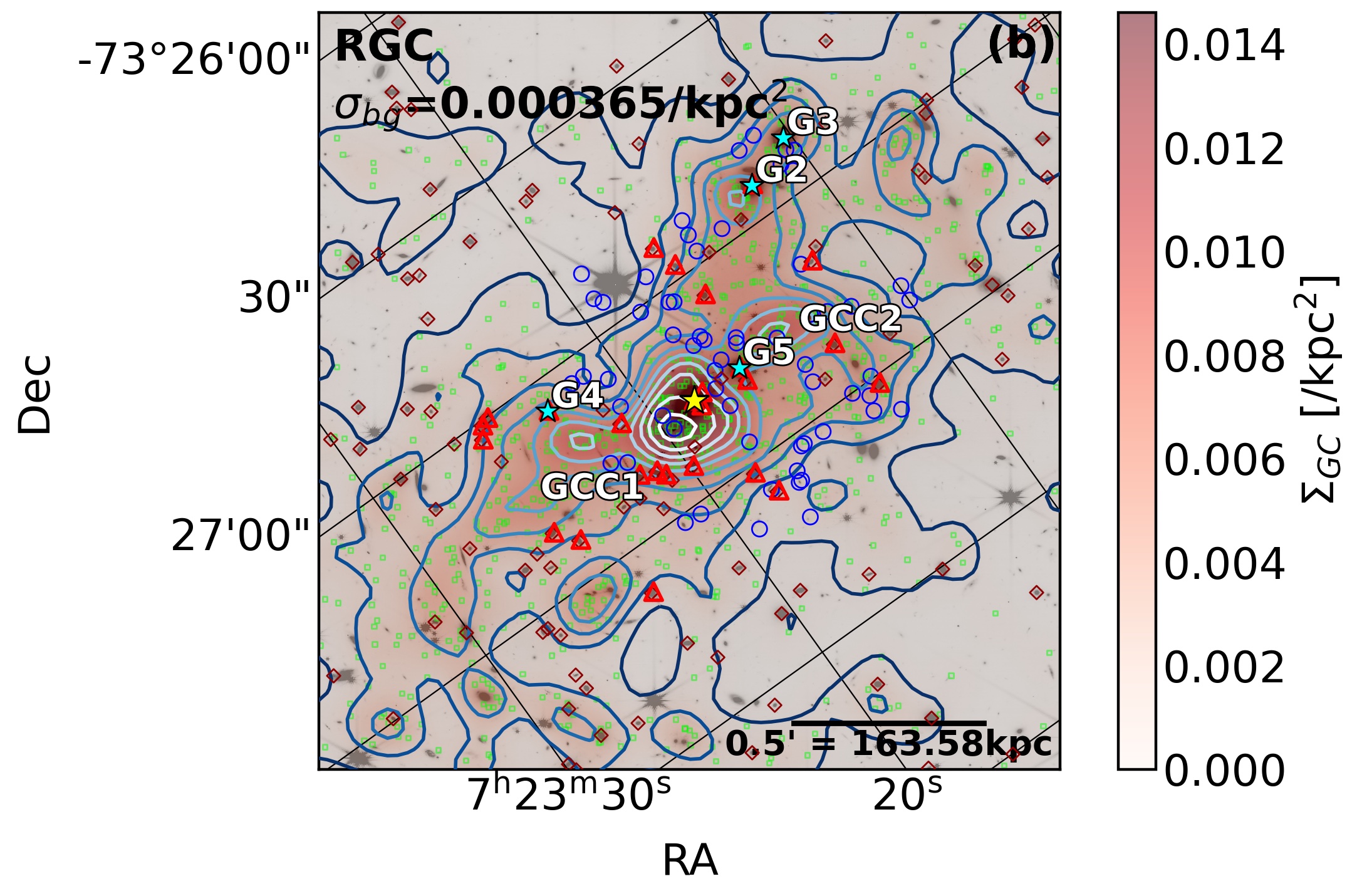} 
    \caption{Same as Figure \ref{fig:GCmap}(b) but for  blue GCs
    ($-0.2<(F150W-F200W)_0 <0.2$) and red GCs ($0.2<(F150W-F200W)_0 <0.6$).
    }
	\label{fig:GCmapBR}
\end{figure*}
\clearpage

\begin{figure*}
    \centering
    \includegraphics[scale=0.25]{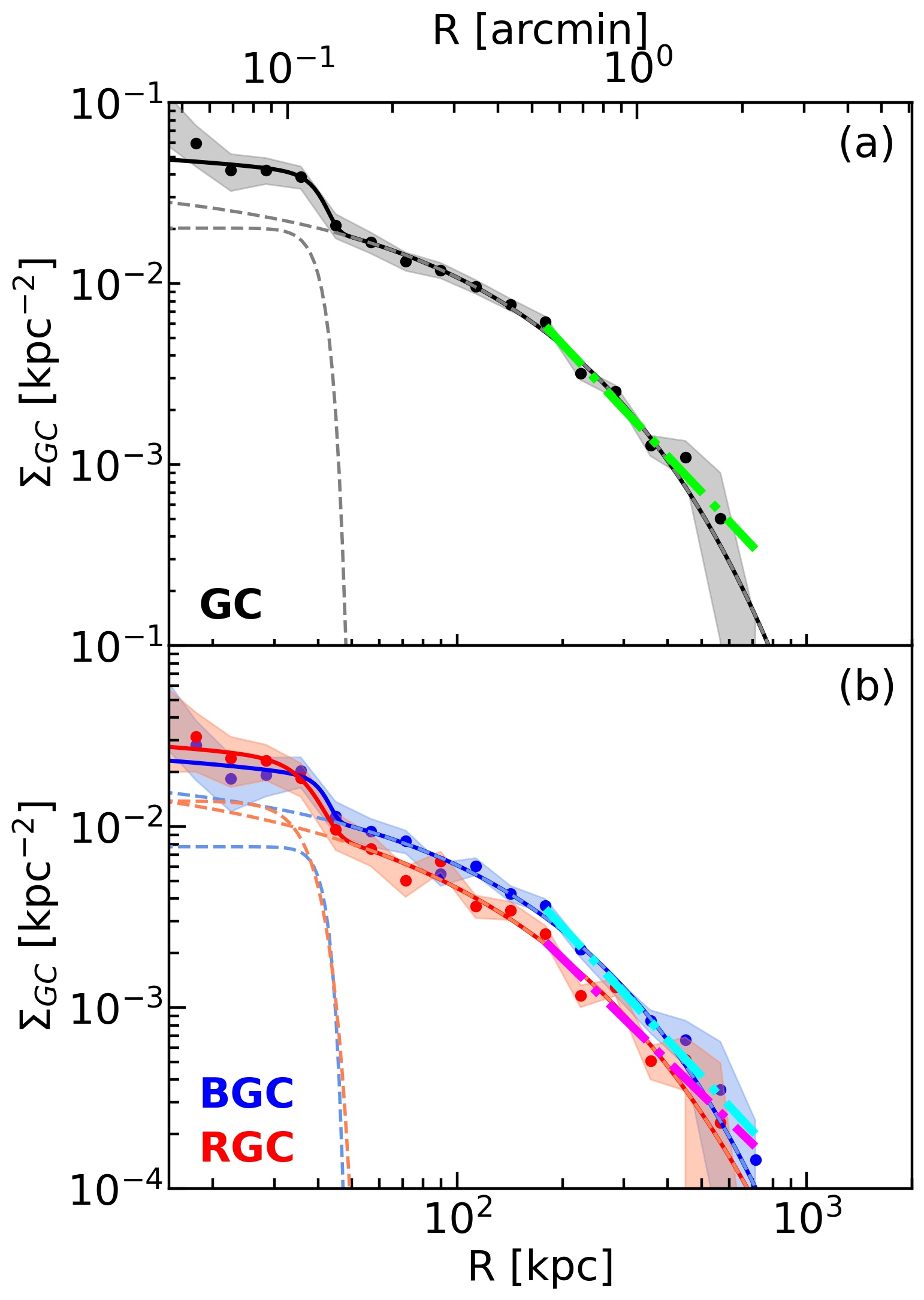}
    \caption{Radial number density profiles of the GCs (all GCs (black circles), blue GCs (blue circles) and red GCs (red circles). Solid  lines represent the two component S\'ersic law fitting results for the central plus intracluster GC components (dashed lines for individual components). The black, blue, and red lines, denote, respectively,  fitting results for the all GCs, blue GCs, and red GCs.
    Thick dot-dashed lines denote the power law fitting results for the outer region ($158<R<631$ kpc). 
    }
	\label{fig:GCRDP}
\end{figure*}
\clearpage

\begin{figure*}
    \centering
\plottwo{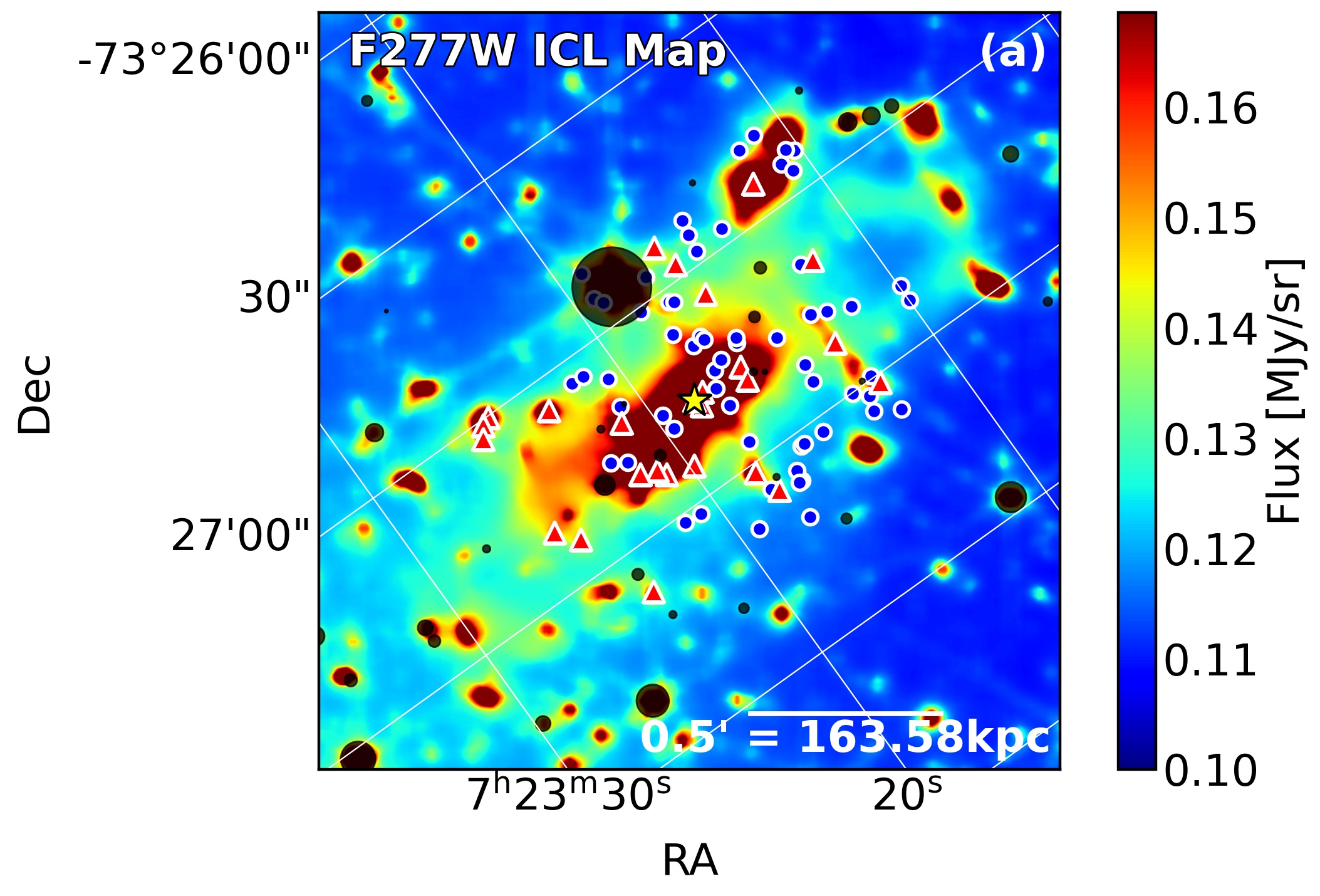}{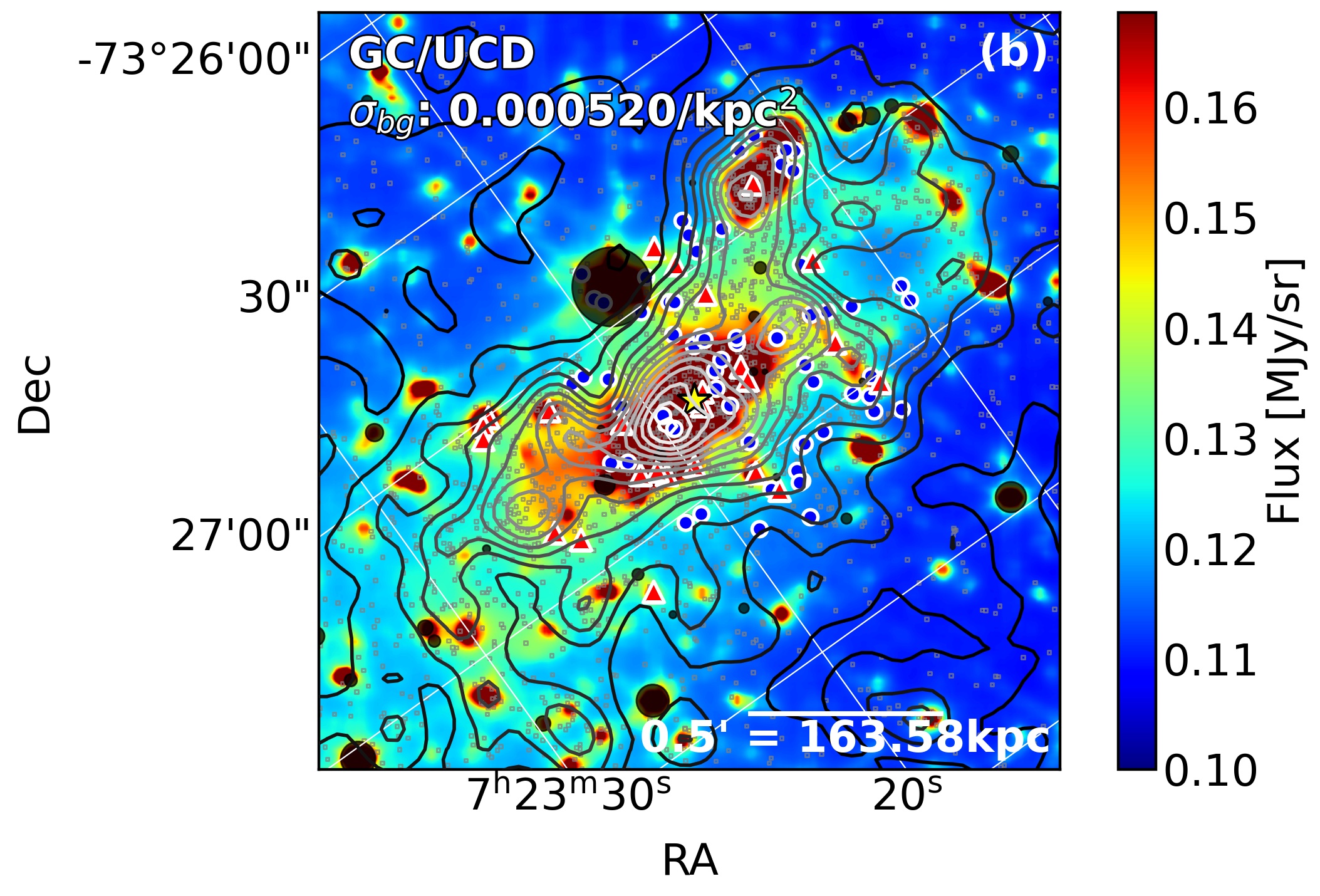} 
\plottwo{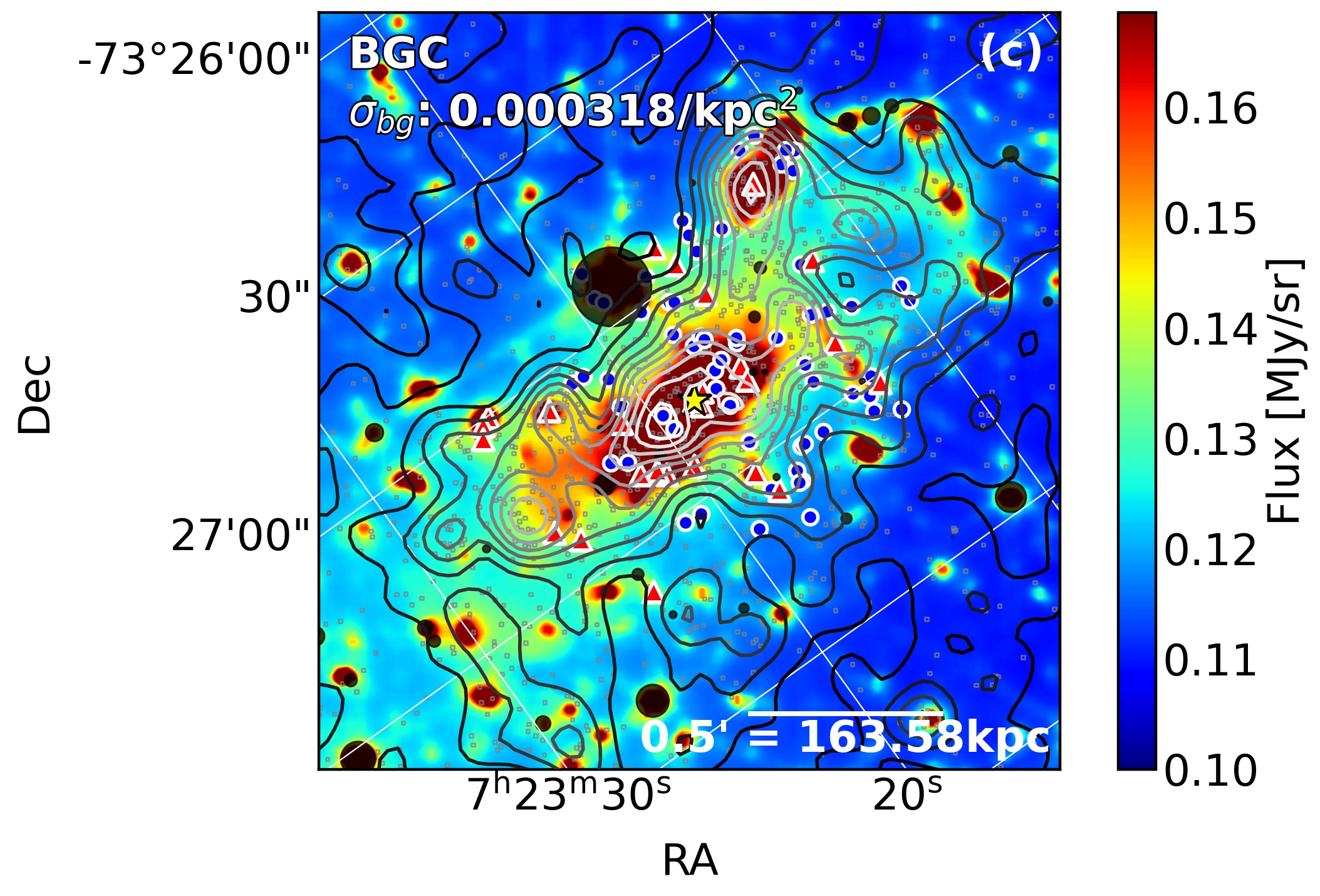}{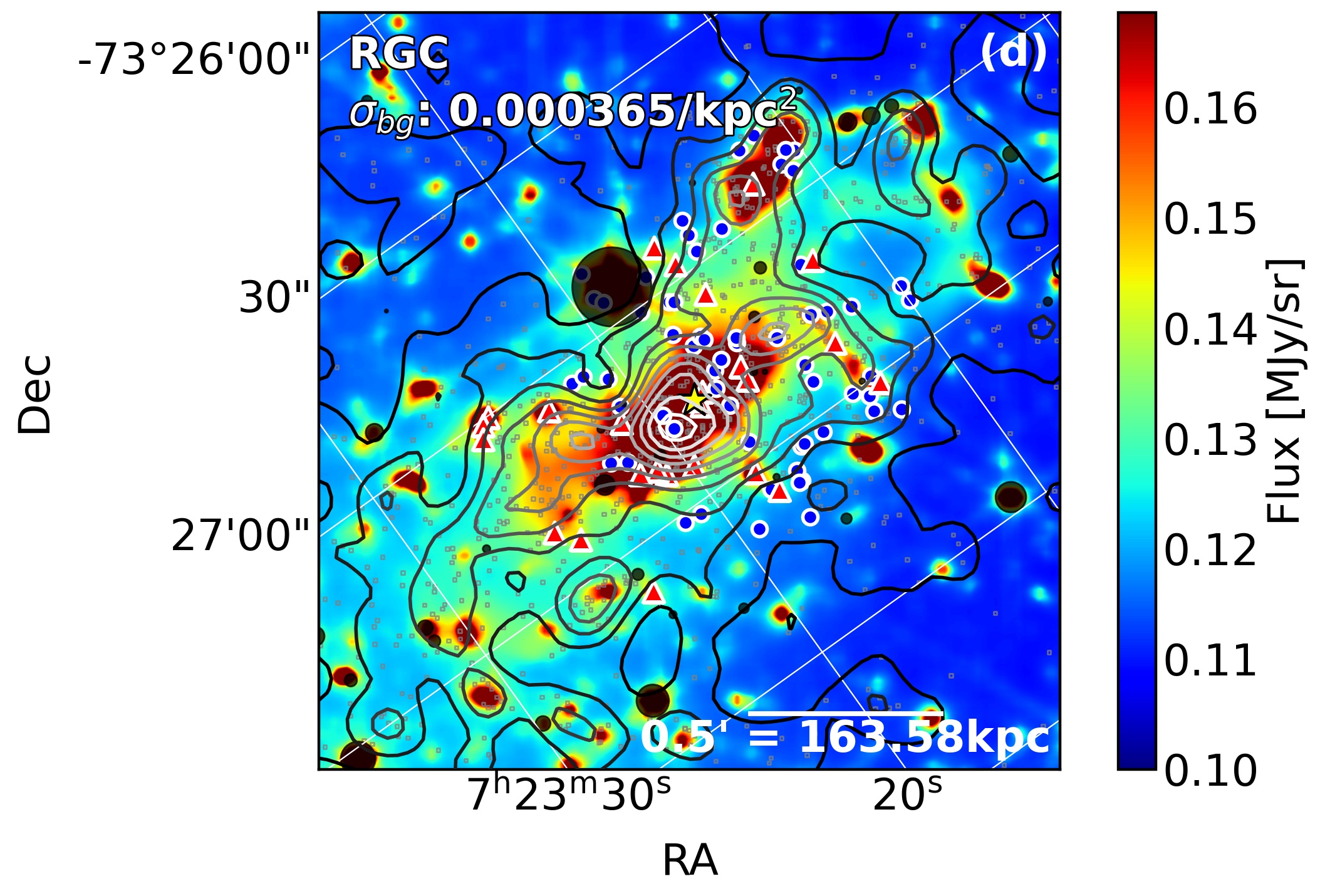} 
    \caption{Comparison of the GC (all GC, blue GC, and red GC) number density contour maps with the ICL map (pseudo color map) of the F277W image of the central field that was derived from  a median  boxcar ($51\times51$ pixels) smoothing filter.
    Red triangles and blue circles represent the known cluster member galaxies and the gravitational lens sources \citep{mah22}, respectively. Black circles denote the bright stars with $F200W_0 <22$ mag, the sizes of which are scaled according to relative magnitudes.
    The yellow starlet mark is the center of the BCG.
    The color scale bars represent
    the ICL surface brightness.}
	\label{fig:ICLmap}
\end{figure*}

\begin{figure*}
    \centering
    \plottwo{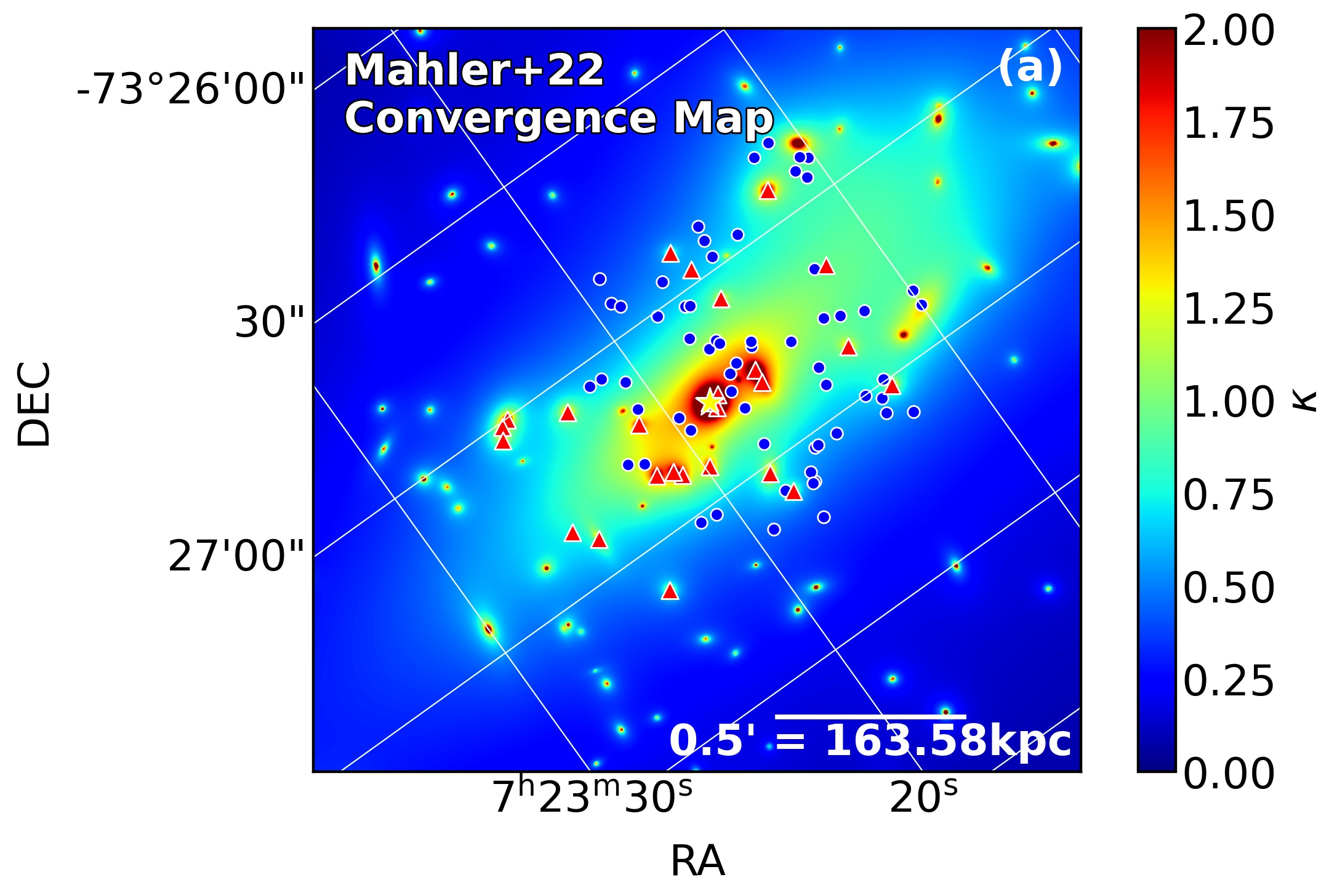}{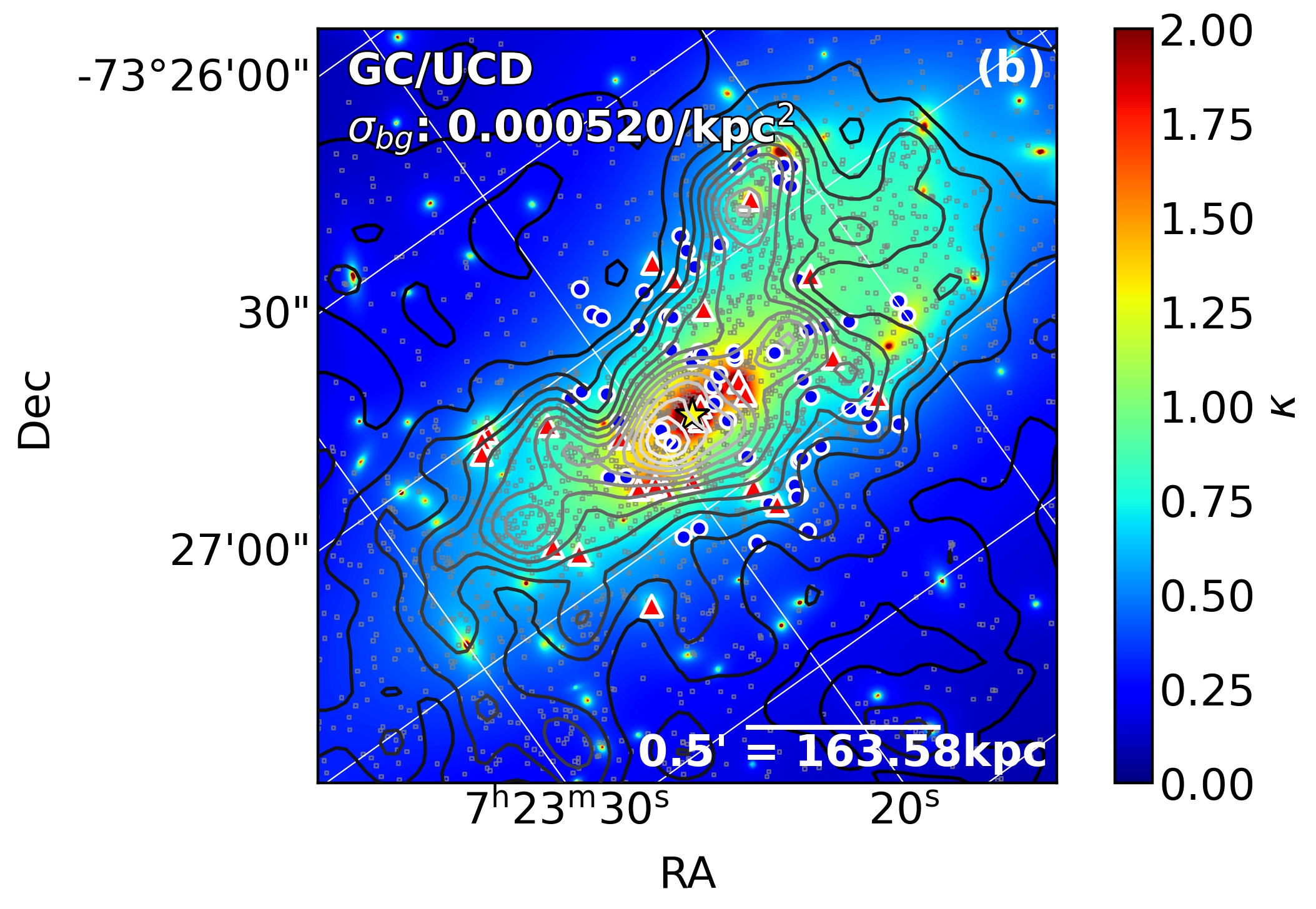} 
    \plottwo{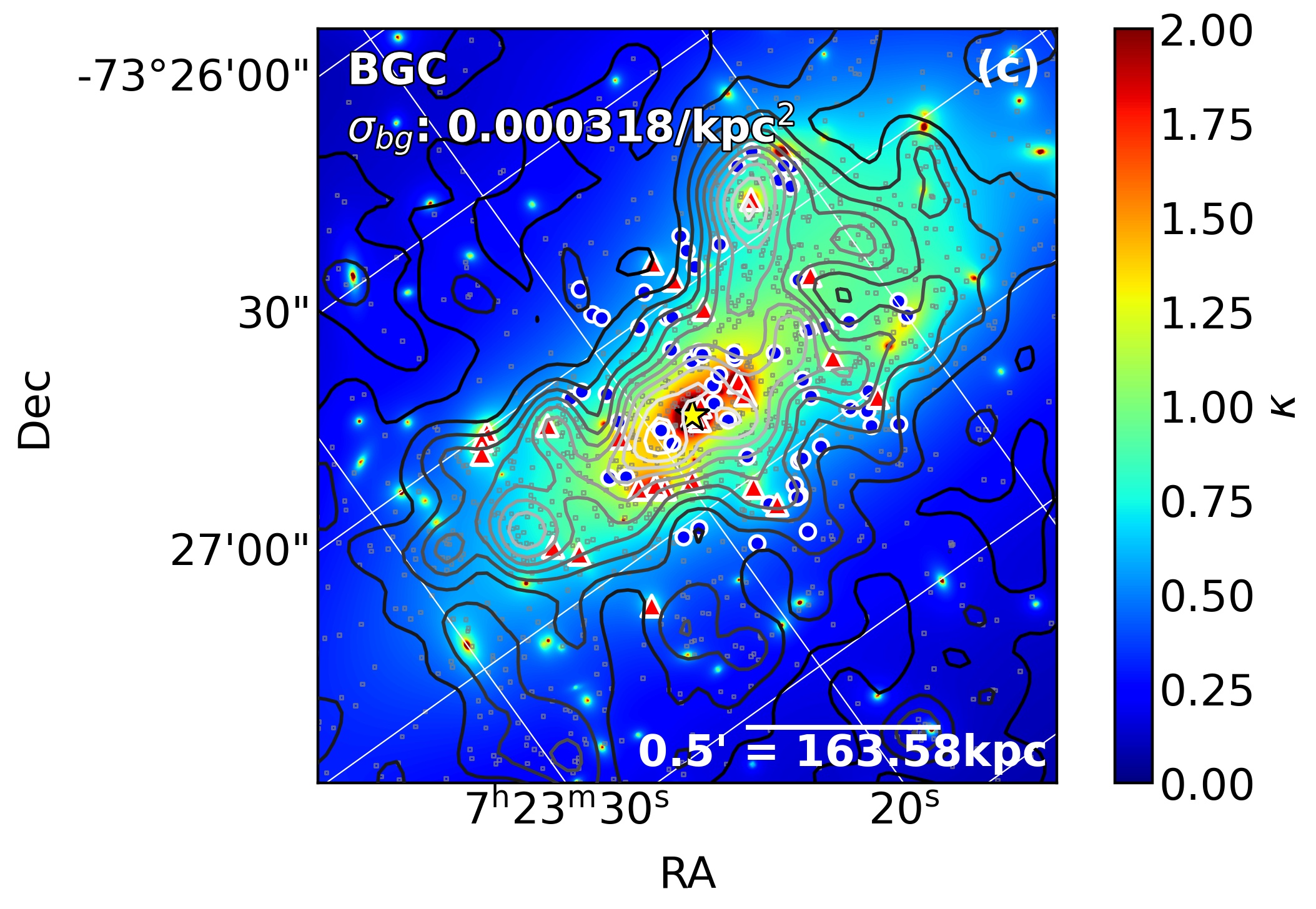}{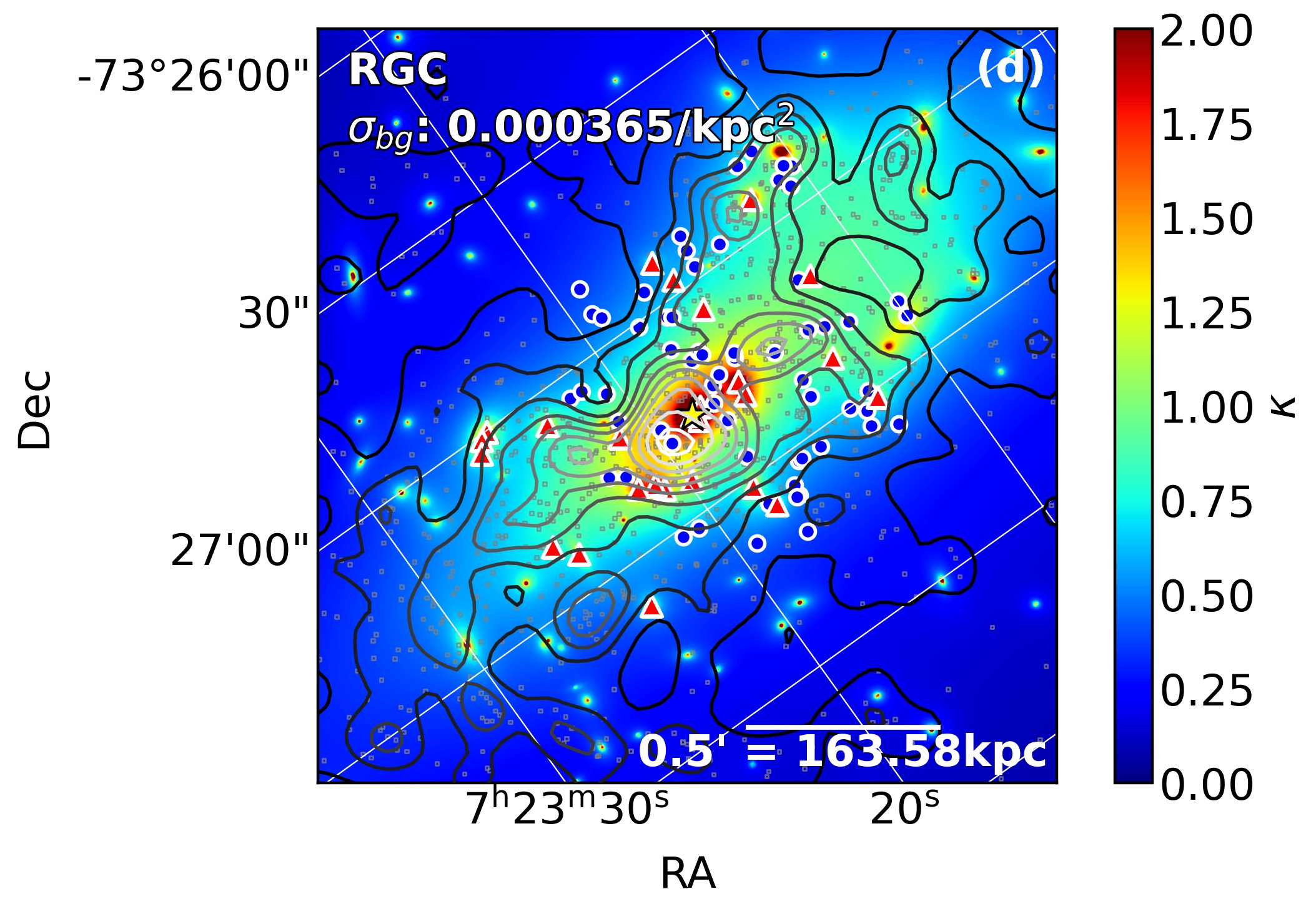}
    \caption{ 
    Comparison of the GC  number density contour maps for the central field of SMACS 0723 with  the dark matter surface mass density map (the convergence map) from strong lensing models given by \citet{mah22} (pseudo color map). 
    Red triangles and blue circles represent the known cluster member galaxies and the gravitational lens sources \citep{mah22}, respectively.
    The color scale bars denote the convergence (the dark matter surface mass density normalized to the critical surface mass density).
    }
	\label{fig:compmassmap}
\end{figure*}

\begin{figure*}
    \centering
     \includegraphics[scale=0.3]{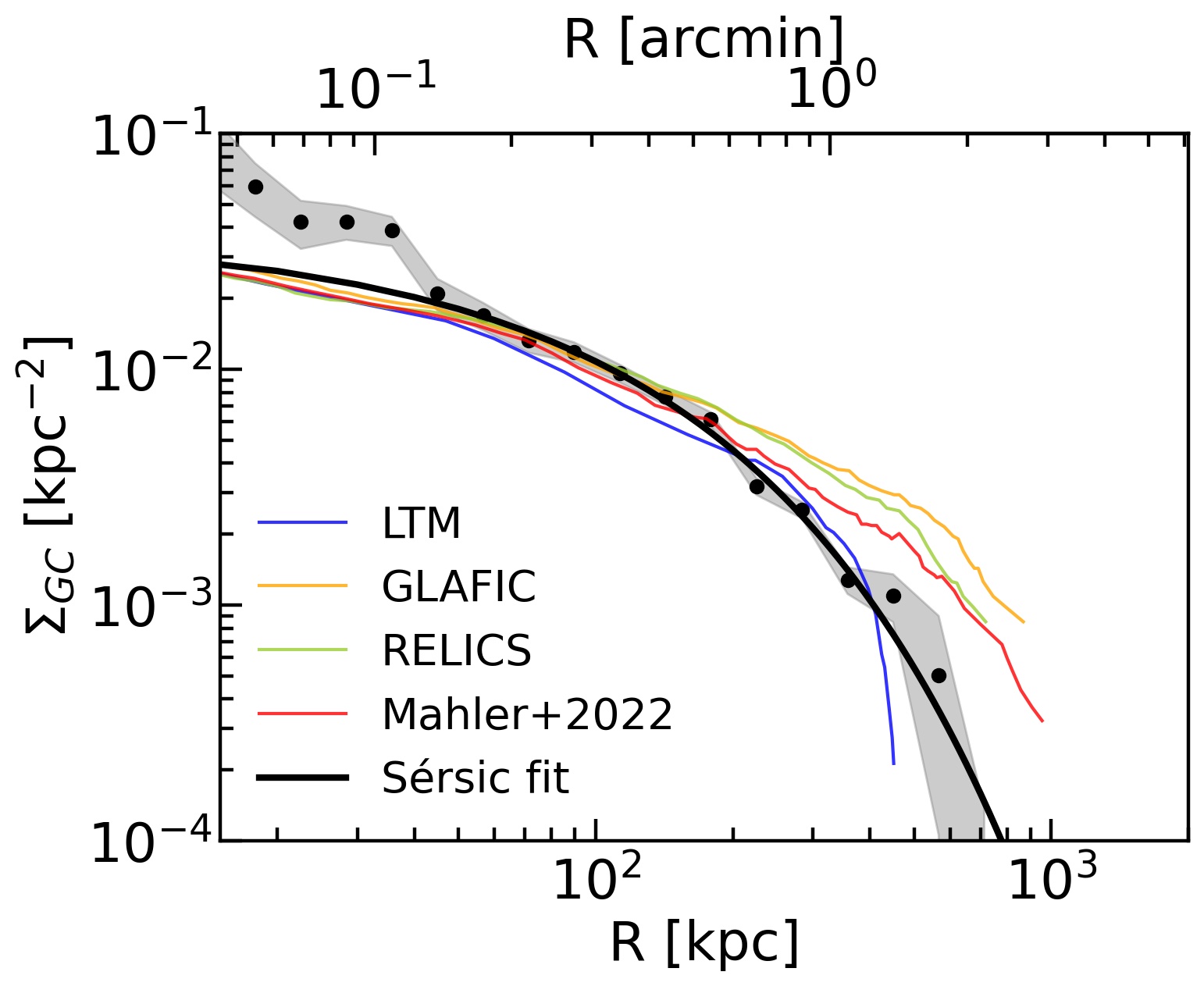} 
    \caption{Comparison of radial number density profiles of the GCs in SMACS 0723 (black circles for the data and the thick black line for the fitting result of the intracluster component) 
    and the dark matter mass density profiles from four strong lensing models given in \citet{mah22} (their Fig. 6): RELICS-lenstool \citep{coe19} (light green line), RELICS-GLAFIC \citep{coe19} (yellow line), LTM \citep{gol22} (blue line), and \citet{mah22} (red line).
    The lensing model profiles were arbitrarily shifted to match the GC data for $50<R<200$ kpc. 
    }
	\label{fig:compRDP}
\end{figure*}
\clearpage
\end{document}